\documentclass[]{tCPH2e}

\usepackage{aas_macros}
\usepackage{color}
\usepackage[utf8]{inputenc}

\begin{document}

\bibliographystyle{JHEP}

\title{A review of indirect searches for particle dark matter}

\author{Jennifer M.~Gaskins$^{\ast,\dagger}$\thanks{$^{\ast}$Einstein Fellow.}\thanks{$^{\dagger}$Current address: GRAPPA, University of Amsterdam, Science Park 904, 1098 XH Amsterdam, Netherlands; email: jennifer.gaskins.astro@gmail.com.}
\\\vspace{6pt}
{\em California Institute of Technology, 1200 E.~California Blvd., Pasadena, California, 91125 USA}
\\\vspace{6pt}\received{\today} }

\maketitle

\begin{abstract}
The indirect detection of dark matter annihilation and decay using observations of photons, charged cosmic rays, and neutrinos offers a promising means of identifying the particle nature of this elusive component of the universe.  The last decade has seen substantial advances in observational data sets, complemented by new insights from numerical simulations, which together have enabled for the first time strong constraints on dark matter particle models, and have revealed several intriguing hints of possible signals.  This review provides an introduction to indirect detection methods and an overview of recent results in the field.
\bigskip
\bigskip

\end{abstract}

\section{Introduction}

\begin{figure}
\centering
\includegraphics[width=0.45\textwidth]{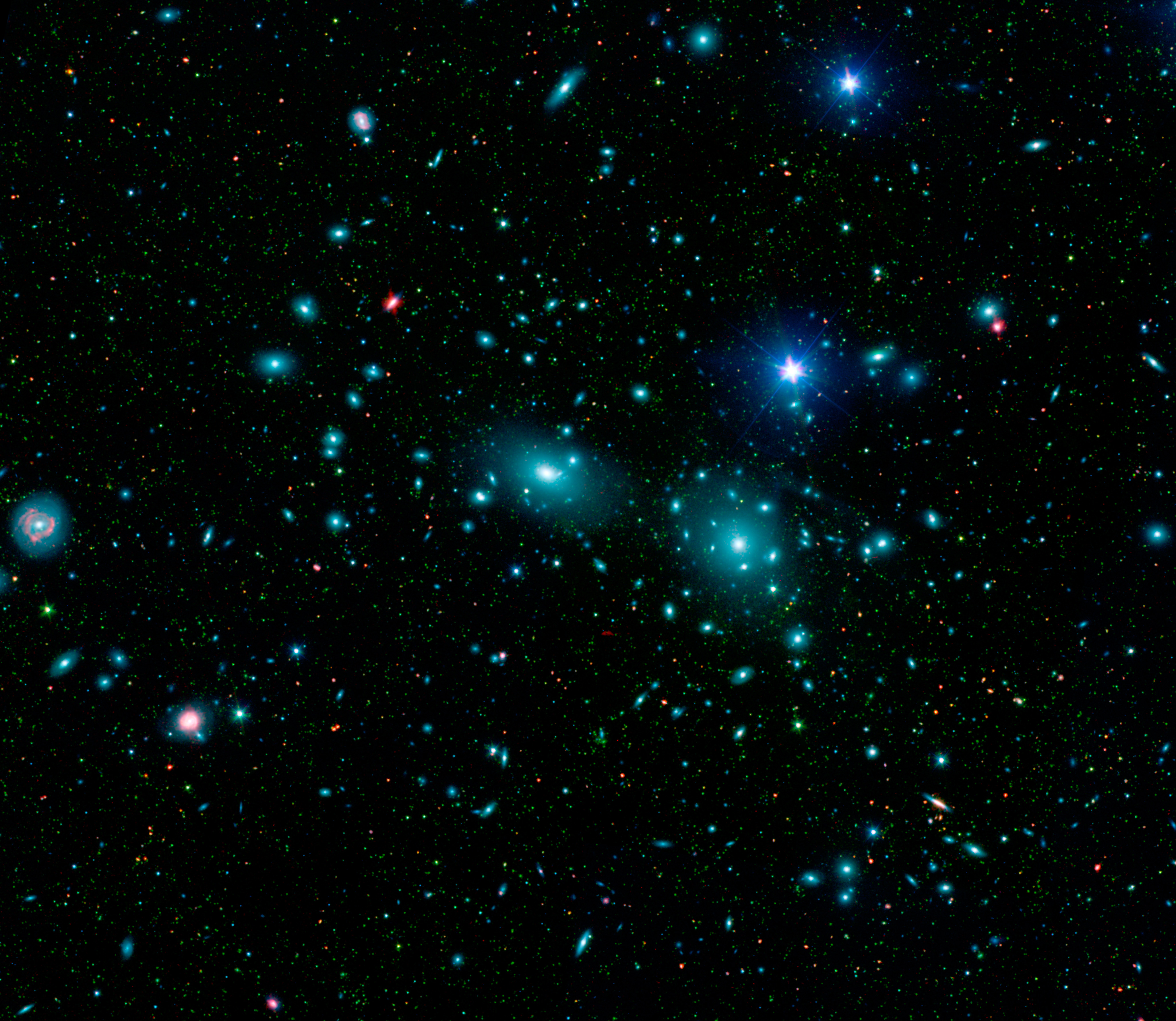}
\caption{The Coma Cluster of galaxies.  False-color mosaic combining images from the Sloan Digital Sky Survey (optical, shown in blue) with long- and short-wavelength infrared images (shown in red and green, respectively) from NASA's Spitzer Space Telescope.  Image credit: NASA / JPL-Caltech / L. Jenkins (GSFC)~\cite{coma,comalicense}.
\label{fig:coma}}  
\end{figure}

More than 80 years ago Swiss astronomer Fritz Zwicky, working at the very same institution where I now write this, observed the motions of galaxies in the Coma Cluster (Fig.~\ref{fig:coma}) from a telescope at the nearby Mount Wilson Observatory.  Zwicky applied the virial theorem to this system and determined that a large amount of invisible matter must be present to keep these galaxies bound together: ``dark matter''~\cite{1933AcHPh...6..110Z}, and with that provided an important clue that what we then knew of the content of the universe was just the tip of the iceberg.\footnote{Earlier suggestions of missing mass from observations of the Milky Way were made by Kapteyn in 1922, also at Mount Wilson Observatory~\cite{Kapteyn:1922zz}, and by Oort in 1932 at Leiden Observatory~\cite{1932BAN.....6..249O}.}

Since that time, an abundance of observations have confirmed the existence of dark matter on a wide range of scales.  Recent results from the \emph{Planck} satellite indicate that dark matter accounts for 83\% of the cosmological matter density~\cite{Ade:2015xua}, and today there is broad consensus that dark matter is a new elementary particle.  However, the identity of the dark matter particle remains one of the outstanding mysteries in modern particle physics, astrophysics, and cosmology.  

This review covers {\em indirect dark matter detection}, a technique that uses astronomical observations of Standard Model (SM) particles to detect the products of the annihilation or decay of dark matter in our Galaxy and throughout the cosmos.  It is distinguished from {\em direct detection}, which aims to detect the scattering of dark matter particles with nuclei in laboratory experiments.  By probing dark matter particle interactions with SM particles, indirect searches are also distinguished from other astronomical observations that investigate only the gravitational interactions of dark matter, such as rotation curve analysis and dark matter studies using gravitational lensing.  Indirect searches complement collider-based searches as well.  Among dark matter detection approaches, indirect searches offer the unique advantage of being able to identify particle dark matter in an astrophysical context, and may one day provide the critical link between a particle detected on Earth at a collider or in a laboratory experiment and the dark matter whose gravitational effects we have observed in the Universe since the 1920s.  Moreover, indirect searches may offer an independent means of mapping the dark matter distribution, yielding insight into the gravitational interplay between dark matter and other components of the universe as well as the particle interactions that can be revealed by understanding the detailed structure of a dark matter halo.  The potential of indirect detection can be further enhanced by leveraging the full complementarity of indirect, direct, and collider approaches.  Global analyses of particle dark matter search results will undoubtedly play an essential role in identifying the dark matter particle and constraining its couplings to SM particles~\cite{Bergstrom:2010gh, Bertone:2011pq, Arina:2013jya, Arrenberg:2013rzp}.  

Many candidate dark matter particles may annihilate and/or decay, and thereby produce indirect signals.  Arguably the most studied category of dark matter candidates is Weakly Interacting Massive Particles (WIMPs), and WIMPs are the focus of most indirect searches.  A wide range of proposed extensions to the SM possess WIMP candidates, such as the lightest supersymmetric particle in Supersymmetry in some scenarios, and the lightest Kaluza-Klein state in theories with universal extra dimensions.  WIMPs characteristically have weak-scale masses, although much more massive candidates termed ``wimpzillas'' have been proposed; their decay may result in ultra-high-energy cosmic rays (UHECRs), motivating indirect searches with those observations.  The right-handed or ``sterile'' neutrino is also a viable dark matter candidate in many scenarios, and can yield indirect signals via its radiative decay to an active neutrino.  

The signals in indirect searches are the SM products of dark matter annihilation and decay.  In practice, these are the stable SM particles that result from an annihilation or decay event; most channels result immediately in unstable SM particles which quickly decay and hadronize into stable states.  Stable states include photons, neutrinos, electrons and positrons, protons and antiprotons, and heavier nuclei and anti-nuclei.  In addition, secondary radiation produced by the subsequent interaction of charged particles originating from annihilation or decay with the environment provides another observational signature.  Searches for dark matter based on its effects on the cosmic microwave background~\cite{Galli:2009zc, Slatyer:2009yq, Galli:2011rz} and the optical depth of the Universe~\cite{Huetsi:2009ex, Cirelli:2009bb} due to energy injection from annihilation during recombination and during and after reionization, can also be very sensitive probes.  Those constraints are not discussed further here, and instead I refer the reader to those works.

Currently there are intensive international efforts to detect these astroparticle messengers and uncover a clear annihilation or decay signal.  We are enjoying an era of abundant data, which has enabled the exploration of new regions of model parameter space, and led to some intriguing hints of possible detections.  Data is providing important input to theory (e.g., claimed signals in experiments spur theorists to build new models to explain the observations), while theoretical work is driving new analyses (e.g., new channels for detection are being identified).  This is an exciting and critical time for indirect searches.  There is a palpable sense of excitement in the community that a robust detection may be just around the corner.

This review aims to provide a concise introduction to indirect detection, including the calculation of indirect signals, targets and techniques for indirect dark matter searches, a summary of the current state of the field, and anticipated future sensitivity.  It is organized as follows: I first give an overview of candidate dark matter particles that may produce signals in indirect searches in~\S\ref{sec:candidates}, followed by a discussion of the dark matter distribution in~\S\ref{sec:dmdistrib}.  \S\ref{sec:signals} presents the calculation of annihilation and decay signals.  The key capabilities of current and planned experiments that can perform indirect searches are surveyed in \S\ref{sec:experiments}.  The following sections focus on specific search targets, observable particles, and recent results in the context of selected dark matter candidates: \S\ref{sec:wimpphotons}, \S\ref{sec:wimpcr}, and \S\ref{sec:wimpnu} cover annihilation and decay signals from WIMPs in photons, cosmic rays, and neutrinos, respectively; UHECRs from superheavy dark matter are briefly reviewed in \S\ref{sec:superheavy}; and sterile neutrino decay signals are presented in \S\ref{sec:sterilenu}.  \S\ref{sec:disc} summarizes the state of indirect dark matter searches, including current results and future prospects.

\section{Dark matter candidate particles for indirect searches}
\label{sec:candidates}

A vast array of candidates for the dark matter particle have been proposed in the context of particle physics models, many of which are expected to produce indirect signals via annihilation or decay, while others (e.g., asymmetric dark matter) generally must be detected by other means.
In this work I consider WIMPs, superheavy dark matter, and sterile neutrinos, since they are expected to annihilate and/or decay to SM particles, and I focus on the detection of their indirect signatures.  For a more detailed discussion of the properties of these and other dark matter candidates, including their particle physics frameworks, I refer the reader to several other excellent reviews~\cite{Jungman:1995df,Bergstrom:2000pn,Bertone:2004pz,Hooper:2007qk}.

\subsection{WIMPs}
\label{sec:wimps}
WIMPs are a broad category of dark matter candidates with a particular set of properties: they couple to the SM via weak interactions, do not directly couple to the photon, and are thermally produced in the early universe with their relic density set by their abundance when they freeze out (i.e., when their interaction rate is sufficiently small that they fall out of equilibrium with other particles).  In the standard freeze-out scenario, the pair annihilation rate of a massive thermal relic particle $\chi$ is directly linked to the relic abundance observed today as~\cite{Jungman:1995df}
\begin{equation}
\label{eq:relicdensity}
\Omega_{\chi}h^{2} \sim \frac{3 \times 10^{-27}\,\, \textrm{cm$^3$ s$^{-1}$}}{\langle \sigma_{\rm A} v \rangle},
\end{equation}
where $\Omega_{\chi}$ is the relic density of the $\chi$ particle in units of the critical density ($\rho_{c}=3H^{2}/8\pi G_{N}$, where $H$ is the Hubble parameter), $h$ is defined via the Hubble constant $H_{0}=h\, 100$~km~s$^{-1}$~Mpc$^{-1}$, $\sigma_{\rm A}$ is the $\chi$ particle's pair annihilation cross section, $v$ is the relative velocity of the $\chi$ particles, and $\langle \rangle$ denotes averaging over the thermal velocity distribution.  In this work $\chi$ will be used to refer to the dark matter particle.  In the indirect detection literature, the quantity $\langle \sigma_{\rm A} v \rangle$ is often referred to as simply the annihilation cross section.  Equation~\ref{eq:relicdensity} refers to the total annihilation cross section, however indirect searches generally consider the cross section for annihilation to a specified final state.  

Based on the measured abundance of dark matter, Eq.~\ref{eq:relicdensity} implies that a particle that constitutes all of the dark matter will have a total pair annihilation cross section of $\langle \sigma_{\rm A} v \rangle \sim 3 \times 10^{-26}$~cm$^3$~s$^{-1}$ (see also the more precise calculation of \cite{Steigman:2012nb}); this value is often used as a benchmark and is referred to as the thermal relic cross section.  This is an appropriate cross-section for a massive particle interacting via the weak force, hence the application of the term WIMP to this type of thermal relic dark matter candidate.  This value is a good choice for a benchmark, however it is important to keep in mind that in some scenarios the annihilation cross section required to produce the observed relic density can vary significantly from the canonical thermal relic value (see, e.g.,~\cite{Bertone:2004pz}). 

The weak-scale masses of WIMPs (tens of GeV to several TeV) imply a similar energy scale for the prompt observable products of annihilation and decay.  Indirect searches for WIMPs are therefore focused largely on gamma rays and high-energy cosmic rays and neutrinos, although searches for secondary emission at lower energy can also be competitive, as discussed in \S\ref{sec:multiwave}.  

The framework of Supersymmetry offers WIMP dark matter candidates, which are extensively reviewed in~\cite{Jungman:1995df}.  Theories of universal extra dimensions also introduce WIMP dark matter candidates, referred to as Kaluza-Klein particles.  Here, however, I take a model-independent approach and consider indirect signatures of generic WIMPs, specifying a model only by the WIMP mass and its annihilation cross sections to different SM final states.

\subsection{Superheavy dark matter}
Superheavy dark matter ($m_{\chi} \gtrsim 10^{12}$~GeV) is an example of a non-thermal relic dark matter candidate, and can be produced in a variety of scenarios, including during or after inflation or through topological defects~\cite{Chung:1998zb}.  These particles have extremely low interaction rates, and are assumed be stable on cosmological timescales, but may annihilate or decay to SM particles which could be detected as UHECRs~(e.g., \cite{Berezinsky:1997hy, Blasi:2001hr, Ibarra:2002rq}).

\subsection{Sterile neutrinos}
The right-handed (or ``sterile'') neutrino $\nu_s$ was proposed by~\cite{Dodelson:1993je} as a dark matter candidate, and it has been shown to be viable as cold, warm, or hot dark matter in different scenarios~(see~\cite{Abazajian:2001nj} and references therein). In general, the neutrino flavor eigenstates ($\nu_\alpha$, with $\alpha=e,\mu,\tau,s$) are a linear combination of mass eigenstates ($\nu_a$, with $a=1,2,...$), and the sterile neutrino has a very small mixing with active neutrinos.  A heavier mass state can radiatively decay to a lighter mass state ($\nu_2 \rightarrow \nu_1 + \gamma$, with $m_2 > m_1$),  and since the sterile neutrino is predominantly composed of $\nu_2$ (in this picture), this is often described as the sterile neutrino decaying to an active neutrino.  This produces a photon line at half of the sterile neutrino mass, which for most viable dark matter candidates is in the keV to MeV energy range.  Line emission provides a means to indirectly detect sterile neutrino dark matter.

\section{The dark matter distribution}
\label{sec:dmdistrib}

The distribution of dark matter is a key input to predicting indirect dark matter signals, and one of the largest uncertainties in those predictions.  Dark matter clusters in halos, which may be triaxial~(e.g., \cite{Bett:2006zy,Navarro:2008kc}) and typically host substructures.  For simplicity, the smooth component of dark matter halos is often modeled as a spherically-symmetric distribution; this is a very good approximation in the central regions of halos.

\begin{figure}
\centering
\includegraphics[width=0.65\textwidth]{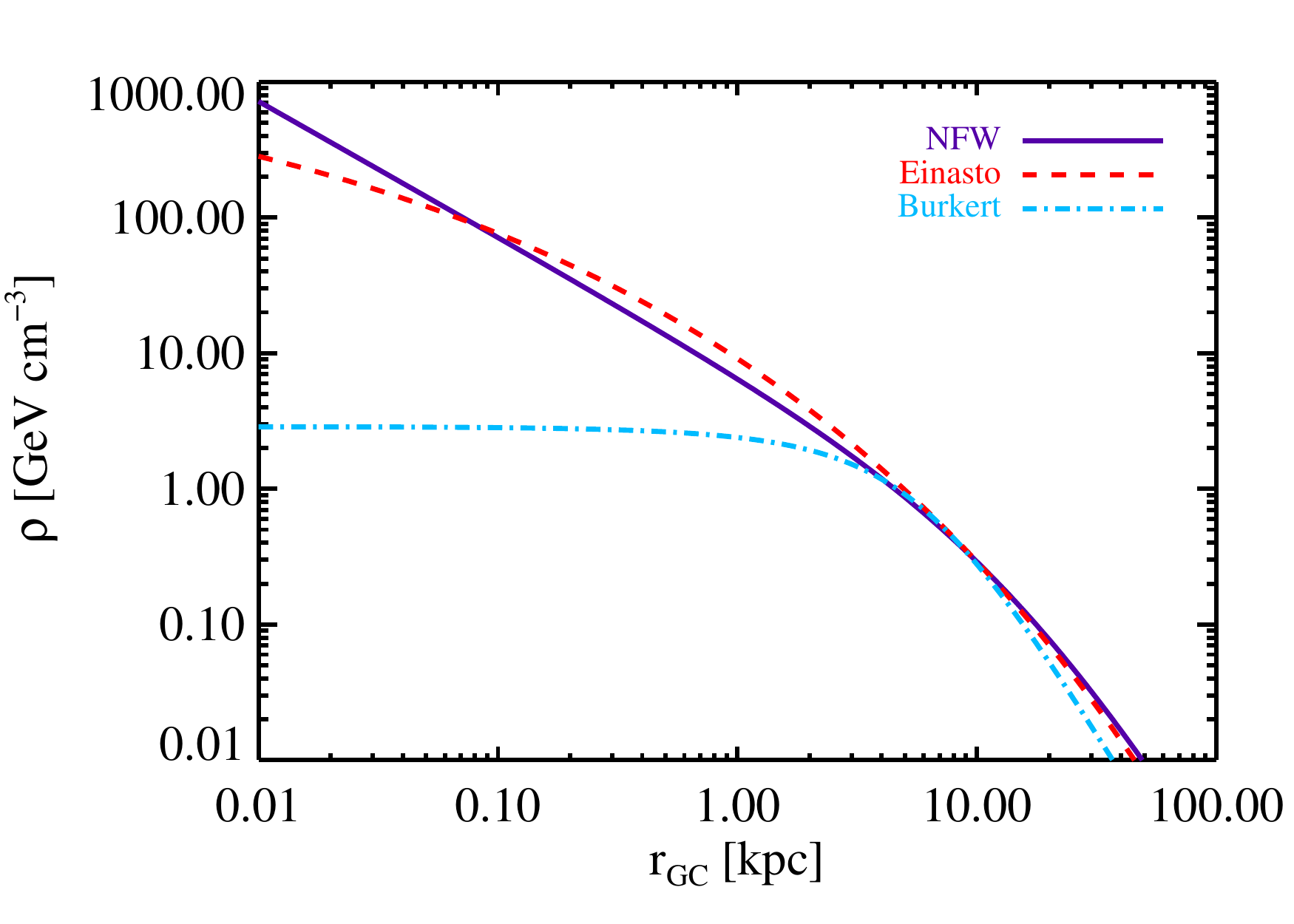}
\caption{Selected Milky Way dark matter halo density profiles.  Profile parameters are mean values obtained by~\cite{Fornasa:2013iaa}; see text for details.\label{fig:profiles}}  
\end{figure}

The dark matter halo density profiles considered today are motivated largely by the results of numerical simulations of structure formation; those simulations historically included dark matter particles only, and did not model baryons.  In the late 1990s simulations showed that a 2-parameter model described the density profile of dark matter halos over a range of halo masses~\cite{Navarro:1995iw,Navarro:1996gj}.  This density profile is referred to as the Navarro-Frenk-White (NFW) profile,
\begin{equation}
\label{eq:NFW}
\rho_{\rm NFW}(r) = \frac{\rho_{0}}{\left(\frac{r}{r_{s}}\right)\left[1+\left(\frac{r}{r_{s}}\right)\right]^{2}},
\end{equation}
where $r$ is the distance from the center of the halo and $r_{s}$ is a scale radius.  For the Milky Way, $r_{s} \sim 20$~kpc~(e.g., \cite{Fornasa:2013iaa}), and the dark matter density at the Sun's position is $\sim 0.4$~GeV/cm$^3$~\cite{Pato:2015dua}.

Some simulation results and observations suggest that the inner slopes of dark matter halos differ from that of the NFW profile, and mechanisms that can modify the inner slope due to interactions with baryons have been proposed.  One such mechanism is adiabatic contraction, which causes the profile to steepen due to the gravitational potential of the baryons pulling in the dark matter~\cite{Gondolo:1999ef, Gnedin:2004cx, Gnedin:2011uj}.  Other mechanisms include feedback from supernovae and interactions between an active galactic nucleus and the interstellar medium which eject gas; these can flatten the inner profile by rapidly modifying the potential, leading to disruption of the dark matter cusp \cite{Mashchenko:2007jp, Pontzen:2011ty, Maccio:2011eh, Governato:2012fa}.  
The NFW profile can be generalized to allow for an arbitrary inner slope $\gamma$,
\begin{equation}
\label{eq:GNFW}
\rho_{\rm GNFW}(r) = \frac{\rho_{0}}{\left(\frac{r}{r_{s}}\right)^{\gamma}\left[1+\left(\frac{r}{r_{s}}\right)\right]^{3-\gamma}}
\end{equation}
where $\gamma=1$ corresponds to the original NFW profile.  The value of $\gamma$ inferred from observations and simulations ranges from $\sim 0$ (a cored profile) to $\sim 1.5$ (as in the Moore profile~\cite{Moore:1999gc}); see~\cite{Adams:2014bda} and references therein.

It has been noted in more recent simulations that the dark matter density profile in the innermost regions of the halo shows deviations from a simple power law, and that a better fit is achieved with a slope that varies with radius~\cite{Navarro:2003ew,Navarro:2008kc,Gao:2007gh}, such as in the profile proposed by Einasto~\cite{1965TrAlm...5...87E},
\begin{equation}
\label{eq:ein}
\rho_{\rm Ein}(r) = \rho_{0} \exp{ \left\{ -\left(\frac{2}{a}\right) \left[ \left( \frac{r}{r_{s}} \right)^{a}-1 \right]\right\} }.
\end{equation}
This profile introduces an extra shape parameter $\alpha$ with respect to the standard NFW profile. For the Milky Way, $\alpha \sim 0.2$~\cite{Iocco:2011jz,Fornasa:2013iaa}.  The scale radius $r_{s}$ is similar to the NFW case for the Milky Way~\cite{Fornasa:2013iaa}.

Observations of dwarf spheroidal~\cite{Walker:2011zu} and low-surface-brightness~\cite{deBlok:2002tg} galaxies have found that some objects are better described by flatter density profiles, and are consistent with the dark matter profile having a central core.  The Burkert profile~\cite{1995ApJ...447L..25B} is an example of a cored profile,
\begin{equation}
\label{eq:burk}
\rho_{\rm Burk}(r) = \frac{\rho_{0}}{\left(1+\frac{r}{r_{s}}\right) \left(1+\frac{r^2}{r_{s}^2}\right)}.
\end{equation}
The Burkert profile exhibits constant density for radii much smaller than the scale radius $r_{s}$.  For the Milky Way $r_{s} \sim 6$~kpc for this profile~\cite{Fornasa:2013iaa}.

The NFW, Einasto, and Burkert density profiles are illustrated in Fig.~\ref{fig:profiles} using parameters appropriate for the Milky Way (mean values reported in~\cite{Fornasa:2013iaa}).  The distributions are very similar at radii outside the solar circle, but can differ substantially in the inner regions of the halo, leading to large variations in indirect signals.

\begin{figure}
\centering
\includegraphics[width=0.45\textwidth]{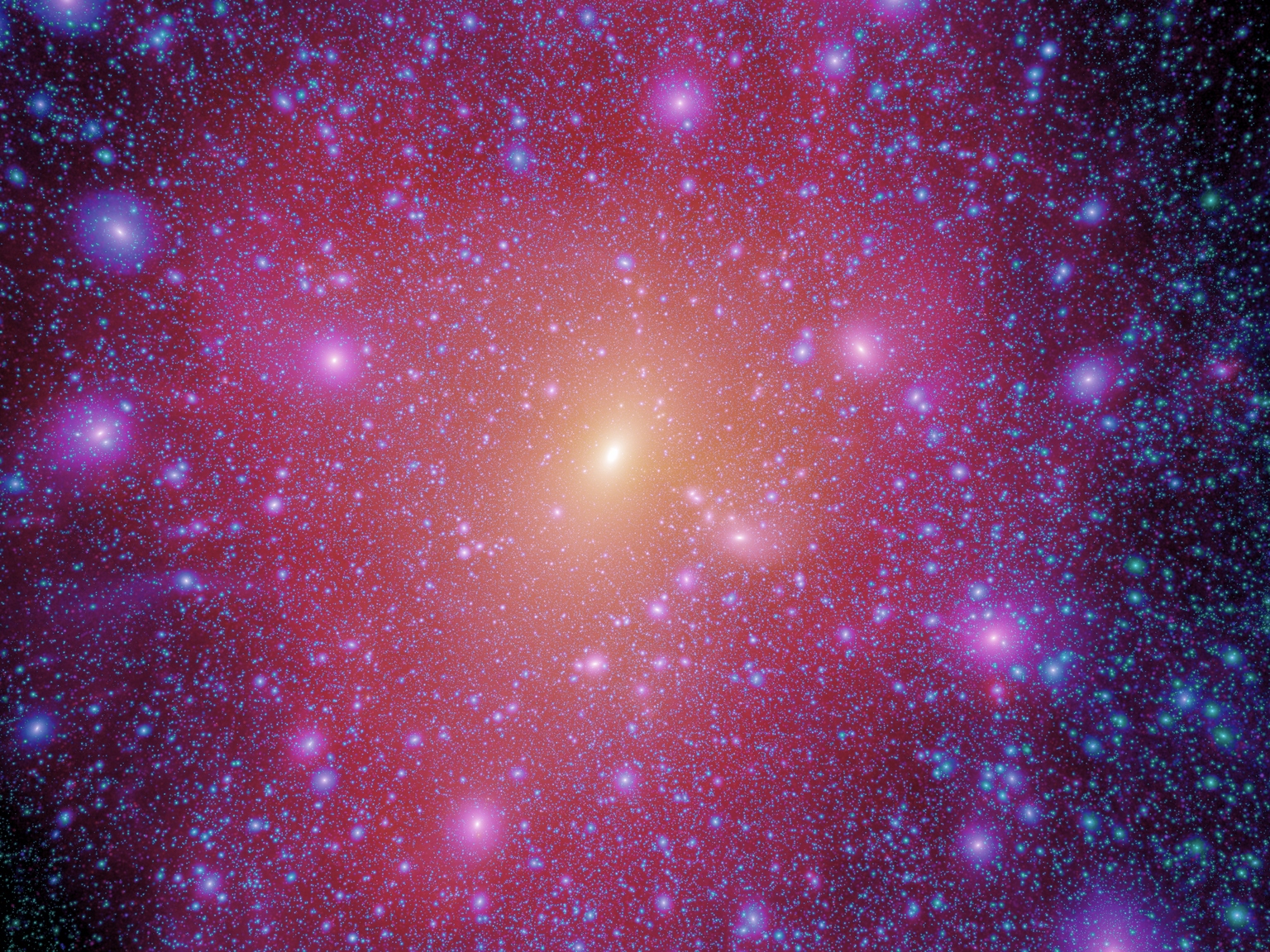}
\caption{Numerical simulation of the dark matter distribution of a galaxy like the Milky Way at the present time, from the Aquarius Project.  The dark matter halo hosts an abundance of subhalos.  The luminous matter would be concentrated in the inner $\sim 10\%$ of the image.  The Aquarius simulations of cold dark matter galactic halos were carried out by the Virgo Consortium~\cite{Springel:2008cc}.\label{fig:aquarius}}  
\end{figure}

Simulations and observations indicate that structure formed hierarchically in the universe, with small halos of dark matter collapsing first, and subsequently merging to form larger objects, although remnants of the original halos typically survive within the merged object.  Consequently, dark matter halos are populated with smaller, denser halos, called subhalos or substructure (Fig.~\ref{fig:aquarius}).  N-body simulations in $\Lambda$CDM cosmologies resolve the high-mass end of the subhalo mass function~\cite{Moore:1999nt, Diemand:2006ik, Springel:2008cc, Stadel:2008pn}, while theoretical arguments suggest that the halo of the Milky Way should be teeming with subhalos of~$\sim 10^{-6}$~M$_\odot$ or smaller~\cite{Bringmann:2006mu,Profumo:2006bv}.  The status of numerical simulations of structure formation including implications for indirect detection is reviewed in~\cite{Kuhlen:2012ft}.

Substructure can have a profound impact on predicted annihilation signals due to the fact that subhalos are denser than the host halo and the rate of annihilation scales as the density squared.  Since the decay signal is directly proportional to mass density, clustering in substructure has an effect only if it modifies the total mass distribution with respect to that of the smooth component alone; in general its effect on halo emission profiles from dark matter decay is negligible except when dealing with individual massive objects within a host halo. 

\section{Annihilation and decay signals}
\label{sec:signals}

\subsection{J-factors}

The prompt flux emitted from the annihilation or decay of dark matter particles can be factored into a part that depends on the particle physics model of the dark matter and a part that is determined by the dark matter distribution.  The latter is referred to as the J-factor, defined as
\begin{equation}
J_{\rm ann}(\psi) = \int_{\rm los}{\rho^{2}(\psi,l) dl}
\end{equation}
for annihilation, and
\begin{equation}
J_{\rm dec}(\psi) = \int_{\rm los}{\rho(\psi,l) dl}
\end{equation}
for decay, where $\psi$ is a sky direction, $l$ is a distance along the line-of-sight (los), and $\rho$ is the dark matter density.  Note that the value of the J-factor is sometimes given as the integral of $J(\psi)$ over a specified angular region.

The differential intensity (particles per area, time, solid angle, and energy) observed from the direction $\psi$ is
\begin{equation}
\label{eq:intensityann}
\frac{dN_{\rm ann}}{dA\,dt\,d\Omega\,dE}=\frac{\langle \sigma v \rangle}{2 m_{\chi}^2}\frac{dN_{x}}{dE} \frac{1}{4\pi} J_{\rm ann}(\psi)
\end{equation}
for annihilation, and 
\begin{equation}
\label{eq:intensitydec}
\frac{dN_{\rm dec}}{dA\,dt\,d\Omega\,dE}=\frac{1}{m_{\chi}\,\tau}\frac{dN_{x}}{dE} \frac{1}{4\pi} J_{\rm dec}(\psi)
\end{equation}
for decay.  Here $\tau$ is the lifetime of the dark matter particle and $dN_{x}/dE$ is the differential spectrum of $x$ particles emitted per annihilation or decay.  The factor of 2 in the denominator of Eq.~\ref{eq:intensityann} applies to dark matter which is its own antiparticle, and becomes a factor of 4 if the dark matter is not its own antiparticle.

\begin{table}
\caption{Approximate J-factors for selected targets, integrated over a circular region with angular radius of 0.5$^{\circ}$, given as $\log_{10}(J_{\rm ann})$ with $J_{\rm ann}$ in GeV$^2$ cm$^{-5}$ sr.  Values obtained from models in~\cite{Fornasa:2013iaa, Geringer-Sameth:2014yza, SanchezConde:2011ap}.\label{tab:jfactors}}\vspace{0.5cm}
\centering
\begin{tabular}{cc}
Target & $\log_{10}(J_{\rm ann})$ \\
\hline \hline
Galactic Center & 21.5\\
Dwarf galaxies (best) & 19\\
Galaxy clusters (best) & 18\\
\hline
\end{tabular}
\end{table}

The most favorable targets for indirect searches are generally those that are relatively nearby, have high dark matter densities, and low backgrounds.  Table~\ref{tab:jfactors} summarizes J-factors for selected targets.  While the Galactic Center has the largest J-factor, strong backgrounds can be a disadvantage for a dark matter search.  Satellite galaxies tend to provide cleaner targets, and combined analysis of multiple satellites can help compensate for the lower J-factor of each individual satellite.  Clusters of galaxies appear to be less optimistic targets, however the uncertainty on the J-factor due to substructure is quite large and substructure could significantly enhance the signal at large radii from the cluster center.
Milky Way dark matter subhalos that do not host a luminous component could be detected either as individual sources, or their collective signal could contribute to measured diffuse emission.  Other targets that have been considered for indirect searches include nearby external galaxies such as M31 and the cumulative signal from cosmological dark matter.

\subsection{Particle spectra}

The total spectrum of $x$ particles emitted per annihilation/decay ($dN_{x}/dE$) can be written as the sum of the spectra produced for all possible final states $f$,
\begin{equation}
\frac{dN_{x}}{dE}=\sum_{f} B_{f}\frac{dN_{x,f}}{dE}
\end{equation}
where $B_{f}$ is the branching ratio to final state $f$ and $dN_{x,f}/dE$ is the spectrum of $x$ particles produced for final state $f$.  The branching ratios to different final states are model-dependent.  Indirect searches often taken a model-independent approach and instead frame results in terms of sensitivity to annihilation or decay to a particular final state.

The final state can be any SM particle pair which is kinematically accessible.  In theories that introduce new particles other than the dark matter particle, the final state can be new particles which then decay to SM particles.  Note that additional final states are allowed for decay~\cite{PalomaresRuiz:2010pn}.  Many final states are not stable, and quickly decay and hadronize to stable particles: photons, neutrinos, electrons and positrons, protons and antiprotons, and heavier nuclei. Prompt emission refers to the stable SM products produced after hadronization and decay in vacuum.  

\begin{figure}
\centering
\includegraphics[width=0.45\textwidth]{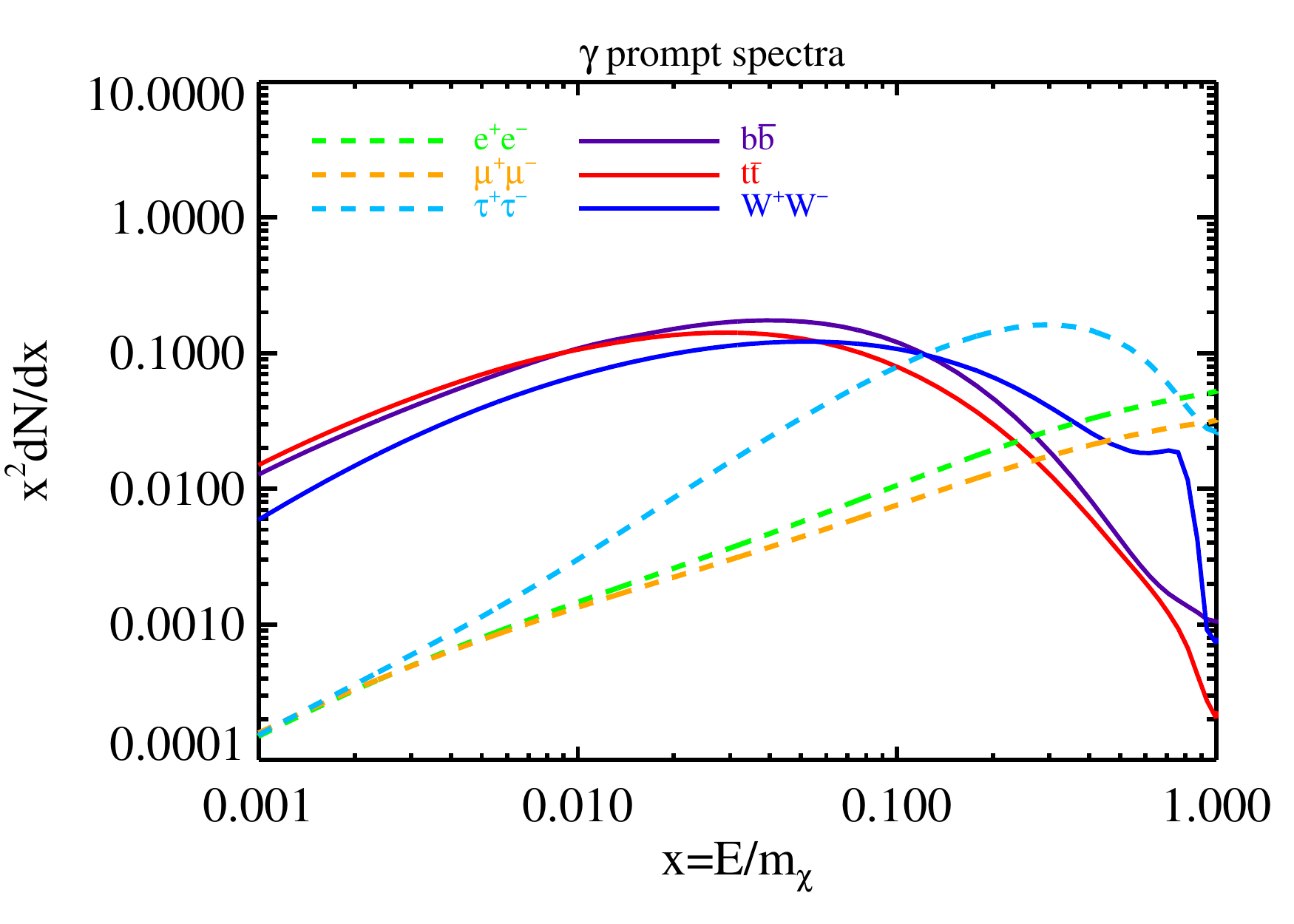}
\includegraphics[width=0.45\textwidth]{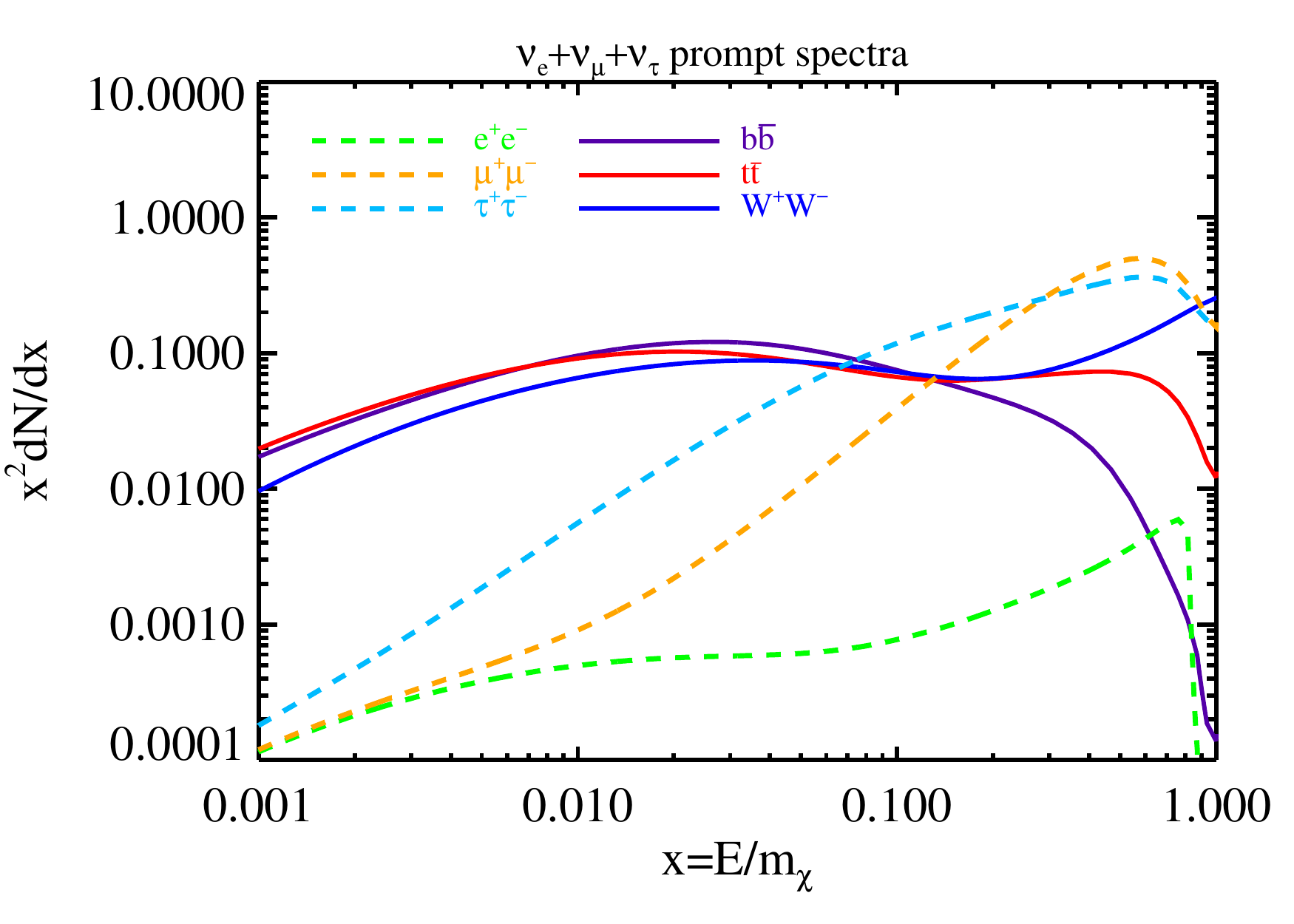}
\vspace{0.5cm}
\includegraphics[width=0.45\textwidth]{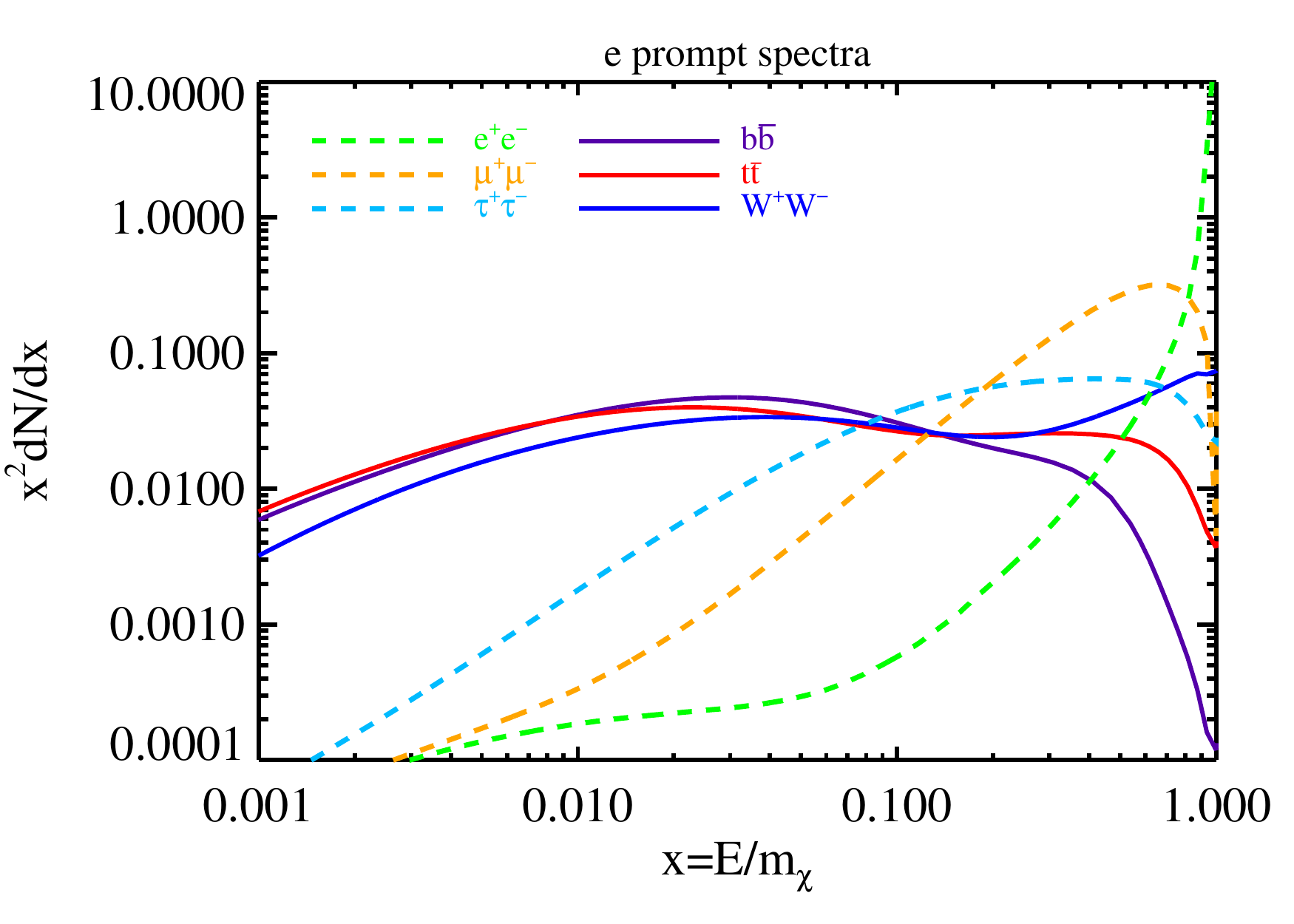}
\includegraphics[width=0.45\textwidth]{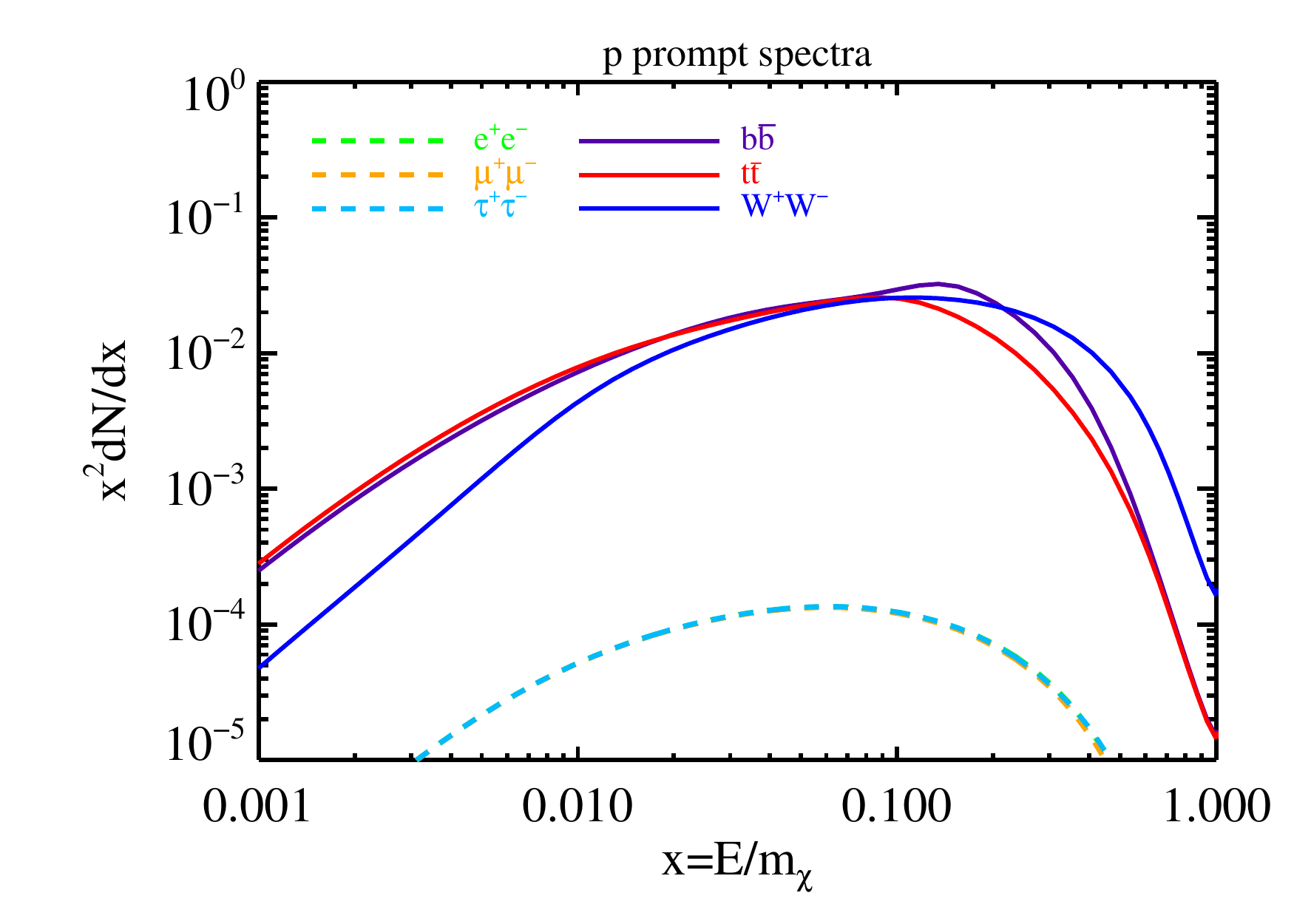}
\caption{Prompt spectra from annihilation of dark matter particles to selected final states for $m_{\chi}= 500$~GeV, for photons (\emph{top left}), sum of all neutrino flavors (\emph{top right}), electrons (\emph{bottom left}), and protons (\emph{bottom right}).  The differential energy spectrum per annihilation is shown in terms of $x=E/m_{\chi}$.  Spectra are valid for decay of a dark matter particle with mass $2m_{\chi}$. Spectra were calculated using PPPC4DMID~\cite{Cirelli:2010xx}.\label{fig:spectra}}  
\end{figure}

The prompt emission spectra of photons, neutrinos, electrons/positrons, and protons/antiprotons associated with several final states are shown in Fig.~\ref{fig:spectra}, calculated using the PPPC4DMID package\footnote{\tt http://www.marcocirelli.net/PPPC4DMID.html}~\cite{Cirelli:2010xx}, which includes electroweak corrections, important for multi-TeV candidates.   Other codes for calculating annihilation spectra have been developed, including the comprehensive, publicly-available software package DarkSUSY~\cite{Gondolo:2004sc}.  Fitting functions for photon spectra are provided in~\cite{Cembranos:2010dm}.

The photon spectra separate fairly cleanly into so-called ``soft'' channels -- quark and gauge-boson final states (here $b\bar{b}$, $t\bar{t}$, and $W^{+}W^{-}$ are shown as representative cases) -- which yield photons largely though the decay of neutral pions produced in hadronization, and ``hard'' channels -- $e^{+}e^{-}$, $\mu^{+}\mu^{-}$, and $\tau^{+}\tau^{-}$ -- which generate photons primarily through final state radiation, which scales as $\sim E^{-1}$.  The $\tau^{+}\tau^{-}$ channel also decays hadronically to pions and produces photons through pion decays, which is the origin of the additional emission for this channel relative to the muon and electron channels.  

Not shown in the figure is photon line emission from $\gamma\gamma$, $Z\gamma$, or $h\gamma$ final states.  At gamma-ray energies, standard astrophysical processes are not known to produce monochromatic emission, making a line detection a smoking gun, unambiguously identifying the origin of the signal as dark matter~\cite{Bergstrom:1989jr, Bergstrom:1994mg, Bergstrom:1997fh, Ullio:1997ke}. Detection of multiple lines would pinpoint the dark matter mass (modulo a factor of 2, pending the determination of the lines to be from annihilation or decay, and assuming proper identification of the final states associated with the lines). However, since WIMPs do not couple directly to the photon, annihilation or decay to lines is loop-suppressed, and in general the continuum photon flux arising from hadronization and decay of the particles produced by other final states easily overwhelms the line signal.  As a result, line searches are particularly challenging, and benefit greatly from large statistics and excellent energy resolution.  A claimed detection of a line in gamma rays is discussed in~\S\ref{sec:130gev}.

The neutrino spectra are shown as the sum of all flavors.  Mixing changes the ratio of neutrino flavors, however over long baselines they fully mix, and exhibit a flavor ratio close to 1:1:1 for $\nu_e:\nu_{\mu}:\nu_{\tau}$, and this is usually the case for indirect targets, with the exception of the Sun and Earth.  In neutrino-based indirect searches, $W^{+}W^{-}$, $\mu^{+}\mu^{-}$, and $\tau^{+}\tau^{-}$ are considered hard channels whereas $b\bar{b}$ and $t\bar{t}$ are soft channels.  The  $e^{+}e^{-}$ channel produces very little neutrino emission.  

For the cases of charged cosmic rays, the most favorable channels for indirect searches for electrons/positrons are leptonic final states, although hadronic final states can also yield appreciable fluxes.  For protons/antiprotons, quark and gauge boson final states produce significant fluxes while leptonic final states yield negligible signals in almost all scenarios.

The spectra shown refer to the prompt emission from annihilation in vacuum.  The observed spectra can differ significantly from those shown due to interactions of the annihilation products with the environment in which they are injected and during propagation to the observer.  These interactions tend to reprocess the annihilation energy carried by the particles, changing the spectral shape and enhancing the low-energy part of the spectrum at the expense of the high-energy part.  

Some of the energy associated with charged particle final states is usually redirected into photons, as occurs in inverse Compton scattering of ambient photon fields, bremsstrahlung off of interstellar gas, synchrotron emission due to propagation in magnetic fields, and hadronic cosmic-ray interactions with interstellar gas producing neutral pions which decay to photons.  For charged particles, propagation to the Earth results in a softer spectrum than was emitted.   For photons, the end result is a multi-wavelength  spectrum spanning radio to gamma-ray energies that depends on the detailed properties of the environment.  The multi-wavelength nature of indirect photon signals is an advantage in that it offers more opportunities for detecting dark matter and could help constrain the properties of the environment, but at the cost of introducing uncertainty in the predicted signals due to poorly understood aspects of the environment.  

In addition to the injection of photons from cosmic-ray interactions, the photon spectrum can also be modified when high-energy ($\gtrsim$~few tens of GeV) photons traveling cosmological distances pair-produce off the extragalactic background light (EBL; the cosmological UV, optical, and IR backgrounds), attenuating the high-energy end of the spectrum and initiating cascades which generate lower-energy photons.  A good knowledge of the EBL over a large redshift range is necessary to calculate cosmological gamma-ray signals from dark matter.

Careful modeling of propagation effects is essential for accurately predicting and interpreting dark matter signals.  Multi-wavelength studies are discussed in~\S\ref{sec:multiwave} and~\S\ref{sec:cosmomulti}.

\section{Indirect search experiments}
\label{sec:experiments}

This section provides a brief survey of recent, current, and planned experiments with indirect detection capabilities.  Instruments, advantages, and challenges of indirect searches with each astroparticle are collected in Table~\ref{tab:searches}.

\begin{table}
\caption{Astroparticles for indirect searches, experiments, advantages, and challenges.  Experiment names in blue refer to planned experiments.  $^{\dag}$Lower-energy photon signatures are considered in \ref{sec:multiwave}; due to the large number of lower-energy observatories, they are not listed individually here.\label{tab:searches}\vspace{0.5cm}}
\centering
\begin{tabular}
{  p{2.2cm}  p{4.8cm}  p{3.7cm}  p{3.7cm}}
Particle & Experiments & Advantages & Challenges\\
\hline \hline
Gamma-ray$^\dag$ photons & Fermi LAT, \textcolor{blue}{GAMMA-400}, H.E.S.S.(-II), MAGIC, VERITAS, HAWC, \textcolor{blue}{CTA} & point back to sources, spectral signatures & backgrounds, attenuation\\
\hline
Neutrinos & IceCube/DeepCore/\textcolor{blue}{PINGU}, ANTARES/\textcolor{blue}{KM3NET, BAIKAL-GVD},
Super-Kamiokande/\textcolor{blue}{Hyper-Kamiokande} & point back to sources, spectral signatures & backgrounds, low statistics\\
\hline
Cosmic rays & PAMELA, AMS-02, ATIC, IACTs, Fermi LAT, Auger, \textcolor{blue}{CTA, GAPS} &
spectral signatures, low backgrounds for antimatter searches & diffusion, do not point back to sources\\
\hline
\end{tabular}
\end{table}

\subsection{Gamma-ray telescopes}

Gamma rays are an excellent astroparticle for indirect searches for WIMP dark matter.  The mass scale of WIMPs implies that a sizable fraction of the emission generated by annihilation and decay ends up at gamma-ray energies.  Furthermore, gamma rays travel to the observer without deflection, allowing mapping of the sources of the signal, and the prompt emission carries important spectral information that can be used to characterize the dark matter particle in the case of a detection.  Together, spatial and spectral signatures can be extremely useful for understanding dark matter properties through indirect searches~\cite{PalomaresRuiz:2010pn}.  

The Earth's atmosphere is opaque to gamma rays, so to directly detect photons at these energies it is necessary to observe from space.  The Fermi Large Area Telescope (LAT)~\cite{Atwood:2009ez, Ackermann:2012kna}  is the primary instrument on the Fermi Gamma-ray Space Telescope, launched in June 2008.  The LAT is a pair-production detector which consists of an array of modules that form the tracker and calorimeter, surrounded by an anti-coincidence detector for charged particle identification.  Sensitive to gamma rays from $\sim 20$~MeV to more than 300~GeV, the LAT detects and reconstructs individual gamma-ray and charged-particle events, determining the arrival direction and energy of each event.  It features a large field-of-view ($\sim 2.4$~sr) and operates primarily in sky-scanning mode, enabling studies of sources all over the sky, including large-scale diffuse emission.  The GAMMA-400 telescope, with a planned launch in 2019, will cover a similar energy range to the LAT but with improved angular and energy resolution~\cite{Moiseev:2013vfa}.

The flux of gamma rays decreases quickly with increasing energy, so instruments with much larger effective areas than feasible with space-based telescopes like the LAT are necessary to observe at higher energies.  Ground-based imaging atmospheric Cherenkov telescopes (IACTs) detect the Cherenkov light produced by particle showers induced by a gamma ray or cosmic ray interacting in the Earth's atmosphere.  IACTs are sensitive from energies of a few tens of GeV to more than 100 TeV, with effective areas up to $\sim 5$~orders of magnitude greater than that of the LAT, providing enhanced sensitivity at  high energies.  By virtue of the Cherenkov technique, IACTs are also sensitive to cosmic-ray--induced showers.  IACTs can identify and reject hadronic cosmic-ray showers with high efficiency, although electron-induced showers are indistinguishable from those originating from gamma rays, and represent a large, irreducible (but virtually isotropic) background for these instruments.  It is important to note that IACTs have a much smaller field of view than the LAT, of order a few degrees (up to~$\sim 10^{\circ}$ for upcoming instruments), forcing these observatories to choose targets carefully.  The currently-operating generation of IACTs includes H.E.S.S.(-II), MAGIC, and VERITAS; these telescopes have placed the strongest bounds on signals from dark matter with $m_{\chi}$ greater than a few hundred GeV via null searches toward a variety of targets~\cite{Abazajian:2011ak, Aleksic:2013xea, Acciari:2010ab, Aliu:2012ga}.  The planned Cherenkov Telescope Array (CTA) observatory~\cite{Acharya:2013sxa}, which is anticipated to begin observations within a few years, will provide significantly improved sensitivity to indirect dark matter signals for WIMPs with $m_{\chi} \gtrsim 100$~GeV and is expected to be able to probe thermal relic cross sections for a range of WIMP masses and channels~\cite{Doro:2012xx, Wood:2013taa, Pierre:2014tra}.  

The HAWC observatory, a water-Cherenkov detector at high altitude, has recently begun observations. HAWC detects gamma rays above 100~GeV and will have sensitivity to high-mass dark matter candidates~\cite{Abeysekara:2014ffg}.

\subsection{Neutrino detectors}

Neutrinos, like gamma rays, preserve spectral information and point back to the source, making them a useful astroparticle for indirect searches.  Detection of astrophysical neutrinos generally involves instrumenting a large volume of water or ice and detecting the Cherenkov light produced in the detector medium as the products of neutrino interactions pass through it.  Large volumes are needed to amass sufficient statistics for neutrino-based dark matter searches.

The IceCube neutrino observatory~\cite{Achterberg:2006md}, completed in 2010, is a cubic kilometer of ice at the South Pole instrumented with photomultiplier tubes to detect Cherenkov light.  It has an energy threshold of $\sim 100$~GeV.  A denser infill within IceCube called DeepCore~\cite{Collaboration:2011ym} reduces the energy threshold to about~$\sim 10$~GeV, and the planned upgrade PINGU will further reduce the threshold to a few~GeV~\cite{Rott:2012gh}.
ANTARES is a large water Cherenkov neutrino telescope in the Mediterranean Sea with an energy threshold as low as $\sim 20$~GeV \cite{Collaboration:2011nsa}.  A next-generation telescope in the Mediterranean with a cubic kilometer or larger volume has been proposed, called KM3NET~\cite{Katz:2006wv}.
Currently a gigaton-volume detector is under construction at Lake Baikal (BAIKAL-GVD)~\cite{Avrorin:2013sla}, which will also offer sensitivity to dark matter signals.

Super-Kamiokande (Super-K) is a water Cherenkov neutrino detector located underground in the Kamioka mine in Japan.  It has been in operation since 1996 and has an energy threshold of $\sim 5$~MeV~\cite{Fukuda:2002uc}.  A proposed next-generation detector, Hyper-K, would feature a volume larger by a factor of $\sim 20$ and would greatly improve sensitivity~\cite{Abe:2011ts}.

\subsection{Cosmic-ray detectors}

Searches for products of dark matter annihilation and decay in local charged cosmic-ray fluxes can be highly sensitive, especially due to low backgrounds for antimatter produced by standard astrophysical processes.  A major challenge for these searches is that it is difficult to identify the locations of the sources of cosmic rays due to diffusion in the Galaxy.

Several current and planned missions aim to measure the flux of antimatter cosmic rays from space.
PAMELA is a space-based instrument to detect charged cosmic rays~\cite{Picozza:2006nm}. Launched in 2006, it is mounted to the Russian satellite Resurs-DK1\@.  The detector includes a magnetic spectrometer which allows it to measure the charge and sign of detected particles, and is sensitive to particle energies from less than 100~MeV to several hundred GeV.  The Alpha Magnetic Spectrometer (AMS-02) is a cosmic-ray detector installed on the International Space Station that measures cosmic rays and gamma rays from a few hundred MeV to a TeV~\cite{Barao:2004ik}.  With its permanent magnet, it can determine the charge and sign of particles passing through the detector.  GAPS is a planned experiment tailored to detect anti-deuterons~\cite{Fuke:2008zz}, which can be a sensitive probe of dark matter annihilation and decay.

Even without distinguishing the sign of the charge, cosmic-ray flux measurements can  be used as an indirect search tool.  ATIC is a balloon-borne experiment to measure cosmic-ray composition from $\sim 100$~GeV to $\sim 100$~TeV~\cite{Chang:2008aa}; it does not distinguish matter and antimatter.  The Fermi LAT and IACTs also have the capability of measuring cosmic rays, and have generated interesting results for the electron+positron spectrum.  Due to the lack of a magnet, these telescopes generally cannot distinguish the sign of particles, however the LAT has cleverly made use of the geomagnetic field to also measure the positron fraction~\cite{FermiLAT:2011ab}.

The Pierre Auger Observatory (Auger) is a ground-based UHECR detector.  Spanning 3000 km$^{2}$, it uses an array of water-Cherenkov surface detectors to detect particles produced in extensive air showers initiated by cosmic-ray primaries, and several optical telescopes that overlook the array to measure fluorescence created by the excitation of nitrogen in the atmosphere by charged particles in the shower.  Auger is sensitive to primary cosmic rays with energies of $10^{19}$~eV and greater~\cite{Allekotte:2007sf,Abraham:2004dt}, allowing it to search for signatures of superheavy dark matter annihilation and decay.

\section{Photons from WIMP annihilation and decay}
\label{sec:wimpphotons}

\subsection{The Galactic Center}

The Galactic Center is one of the most favorable targets for photon-based searches for dark matter due to its close proximity and large concentration of dark matter, making it the brightest dark matter source in the sky.  However, large backgrounds exist at all energies, and disentangling a putative dark matter signal is extremely challenging.  Searches in the Galactic Center have set strong bounds on dark matter annihilation, and revealed multiple possible detections.

\subsubsection{Constraints from gamma-ray searches}

The prompt photon emission spectra (such as the examples shown in Fig.~\ref{fig:spectra}) are a good description of the predicted emission from the Galactic Center at a few tens of GeV and higher, due to the fact that secondary processes do not significantly modify the spectrum at these energies~(e.g., \cite{Regis:2008ij, Bernal:2010ip, Cirelli:2013mqa}).  This corresponds to the high-energy range of the Fermi LAT and the full range of energies accessible to IACTs.  At lower energies, emission from the interaction of cosmic rays must be considered.  
  
The Galactic diffuse emission is the observed gamma-ray flux originating from cosmic-ray interactions, including emission from inverse Compton scattering, bremsstrahlung, and hadronic interactions with the interstellar medium. It represents the dominant background in this region and its properties are 
somewhat uncertain due to the numerous complex processes which together produce this emission.  Detected sources, including the strong source at Sgr A$^{\star}$, also contribute to the emission in this region.  Finally, unresolved members of known source populations such as pulsars may also provide an important contribution to the observed flux.  Modeling emission from unresolved sources is challenging since it relies heavily on extrapolations based on detected members, which are often a small fraction of the total population.

Despite these challenges, tight bounds on dark matter annihilation in many scenarios have been set using searches for emission from the Galactic Center in gamma rays~(e.g., \cite{Abramowski:2011hc,Hooper:2012sr}).  Using Fermi LAT data, current searches exclude the thermic relic annihilation cross section for masses below a few tens of GeV for many annihilation channels for standard density profiles, indicating that these searches are probing theoretically well-motivated dark matter models.

Many searches for gamma rays from dark matter in the Galactic Center set limits on dark matter signals after attempting to model the Galactic diffuse emission (Ref.~\cite{Gomez-Vargas:2013bea} is an exception), leading to results that are somewhat less robust but perhaps more realistic, in the sense that not modeling this known component that certainly contributes a large fraction of the emission would result in overly weak limits.  Future work to better constrain the astrophysical (non-exotic) contributions to the emission have the potential to enhance the sensitivity of these searches~\cite{Siegal-Gaskins:2013tga}.  Multi-wavelength and multi-messenger analysis may also help, along with new approaches to optimize the data for dark matter searches, e.g., the improved angular resolution achieved in the data set developed in~\cite{Portillo:2014ena} and used in the analysis of~\cite{Daylan:2014rsa}, and now available as an option in the latest (Pass 8) Fermi LAT public data.

Dark matter constraints from the non-detection of a significant excess over the astrophysical expectation have been placed by H.E.S.S. for the inner regions of the Galactic Center (including the Sgr A$^{\star}$ source)~\cite{Aharonian:2006wh} and for the halo region around the dynamical center, excluding the plane within $|b| < 0.3^{\circ}$~\cite{Abramowski:2011hc,Abazajian:2011ak} for masses greater than $\sim 300$~GeV.  MAGIC has observed the Galactic Center~\cite{Albert:2005kh}, however that analysis focused on characterizing the central point source rather than on searching for dark matter signatures.  VERITAS has also imaged the Galactic Center region, and an interpretation of that data in terms of dark matter is forthcoming~\cite{Archer:2014jka}.  Due to the fact that VERITAS observes the Galactic Center at large zenith angle, it has enhanced sensitivity to high-energy photons at the expense of raising the energy threshold.  VERITAS anticipates strong sensitivity to dark matter signals from the Galactic Center for high-mass WIMPs~\cite{Smith:2013tta}.  

Current bounds from IACTs are unable to probe cross sections close to thermal, however natural Sommerfeld enhancement from the exchange of $W$ and $Z$ bosons is relevant at large WIMP masses and can lead to annihilation cross sections that are significantly larger than the thermal relic expectation, so IACTs are in fact very close to testing interesting parameter space.

CTA is expected to achieve very strong sensitivity to dark matter signals from the Galactic Center, covering a wider range of masses than previous IACTs and probing annihilation cross sections more than an order of magnitude smaller than those probed by current-generation instruments~\cite{Doro:2012xx, Wood:2013taa, Pierre:2014tra}.  CTA observations of the Galactic Center will also improve on current sensitivity to dark matter decay, and will be able to test the dark matter interpretation of reported cosmic-ray excesses~\cite{Pierre:2014tra}.

\subsubsection{Multi-wavelength radiation}
\label{sec:multiwave}

Due to the complex environment of the Inner Galaxy, dark matter annihilation and decay are guaranteed to generate lower-energy emission from secondary processes~\cite{Regis:2008ij}.  Cosmic rays, whether produced by dark matter or standard astrophysical sources, interact with ambient photon and magnetic fields and interstellar gas, resulting in secondary photon emission from radio to gamma rays.  The most important processes that generate secondary emission in the Inner Galaxy are:
\begin{enumerate}
\item pion decay: cosmic-ray protons interact with the interstellar gas, creating neutral pions that decay to MeV to GeV gamma rays;
\item bremsstrahlung: cosmic-ray electrons interact with the electromagnetic field of the interstellar gas, producing MeV to GeV gamma rays;
\item inverse Compton scattering: cosmic-ray electrons scatter starlight and the cosmic microwave background (CMB) up to X-ray and gamma-ray energies; and
\item synchrotron radiation: the propagation of cosmic-ray electrons in the strong magnetic field of the Galaxy generates emission from radio to X-rays.
\end{enumerate}

The last three processes are competing energy-loss mechanisms for cosmic-ray electrons.  The dominant energy-loss process at any point in the Galaxy varies depending on the interstellar radiation field (ISRF), the distribution of interstellar gas, and the Galactic magnetic field.  A consequence of this complex dependence of the dominant process on the environment is that the spectrum of the multi-wavelength emission associated with a dark matter model can vary significantly throughout the Galaxy~\cite{Cirelli:2013mqa,Tavakoli:2013zva}.

At the Galactic Center, the large magnetic field means that synchrotron is the primary energy-loss process for energetic electrons, and as such it places the strongest constraints from secondary emission at the Galactic Center.  The magnetic field at the Galactic Center is uncertain, however in scenarios with strong magnetic fields, especially those with cuspy or adiabatically-contracted dark matter distributions,  very strong bounds can be placed on WIMP annihilation signals by constraining synchrotron emission using radio data~\cite{Bertone:2002je,Aloisio:2004hy, Regis:2008ij, Bertone:2008xr, Crocker:2010gy, Laha:2012fg}.  X-ray observations can also constrain the synchrotron emission from dark matter at the Galactic Center~\cite{Bergstrom:2006ny}.

\subsubsection{The GeV excess}
\label{sec:gevexcess}

The presence of large backgrounds in the Inner Galaxy make disentangling a dark matter signal and declaring a significant detection extremely challenging~\cite{Siegal-Gaskins:2013tga}, but for one claimed detection it has proved difficult to exclude the possibility of a dark matter origin.

\begin{figure}
\centering
\includegraphics[width=0.45\textwidth]{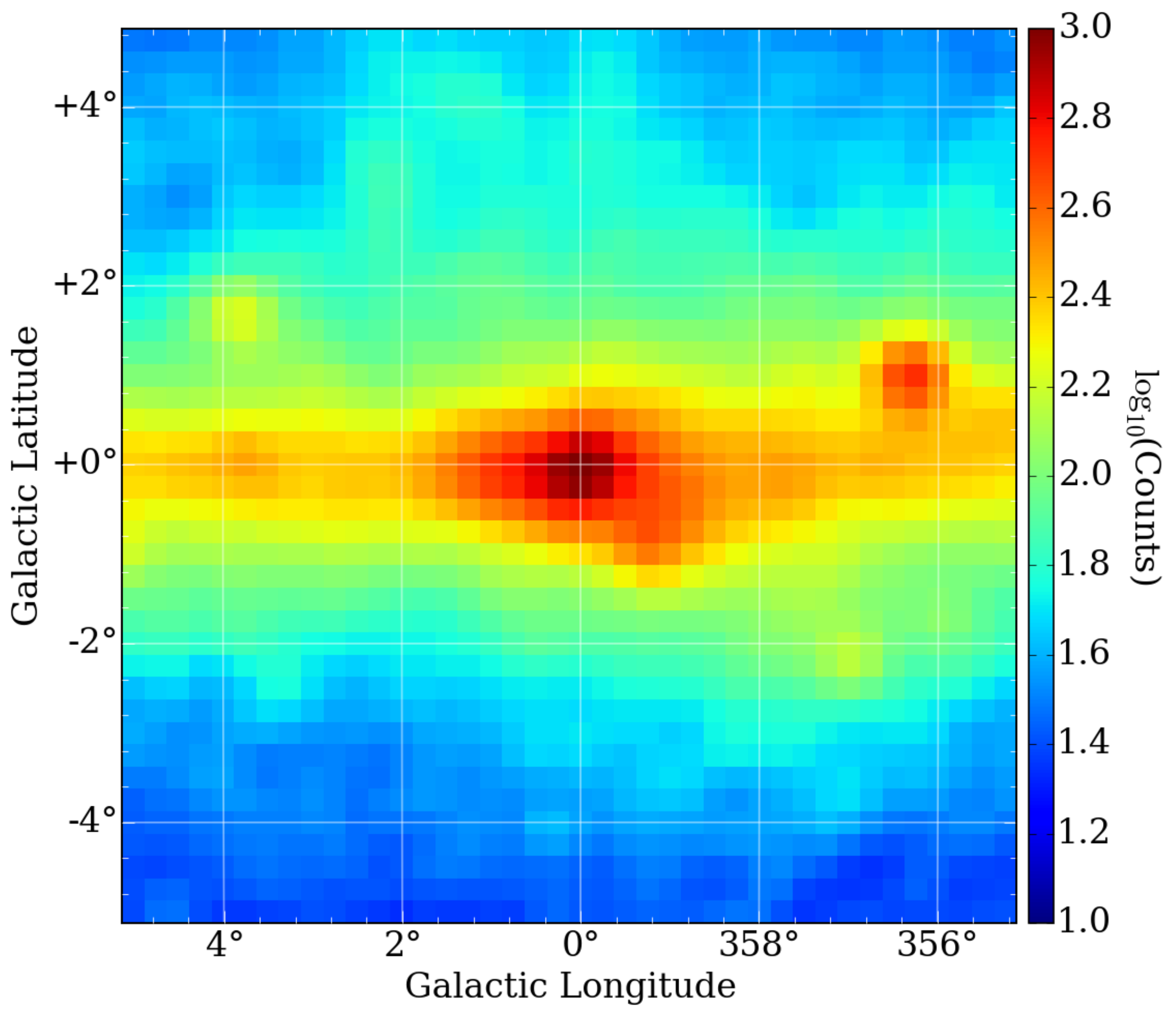}
\includegraphics[width=0.45\textwidth]{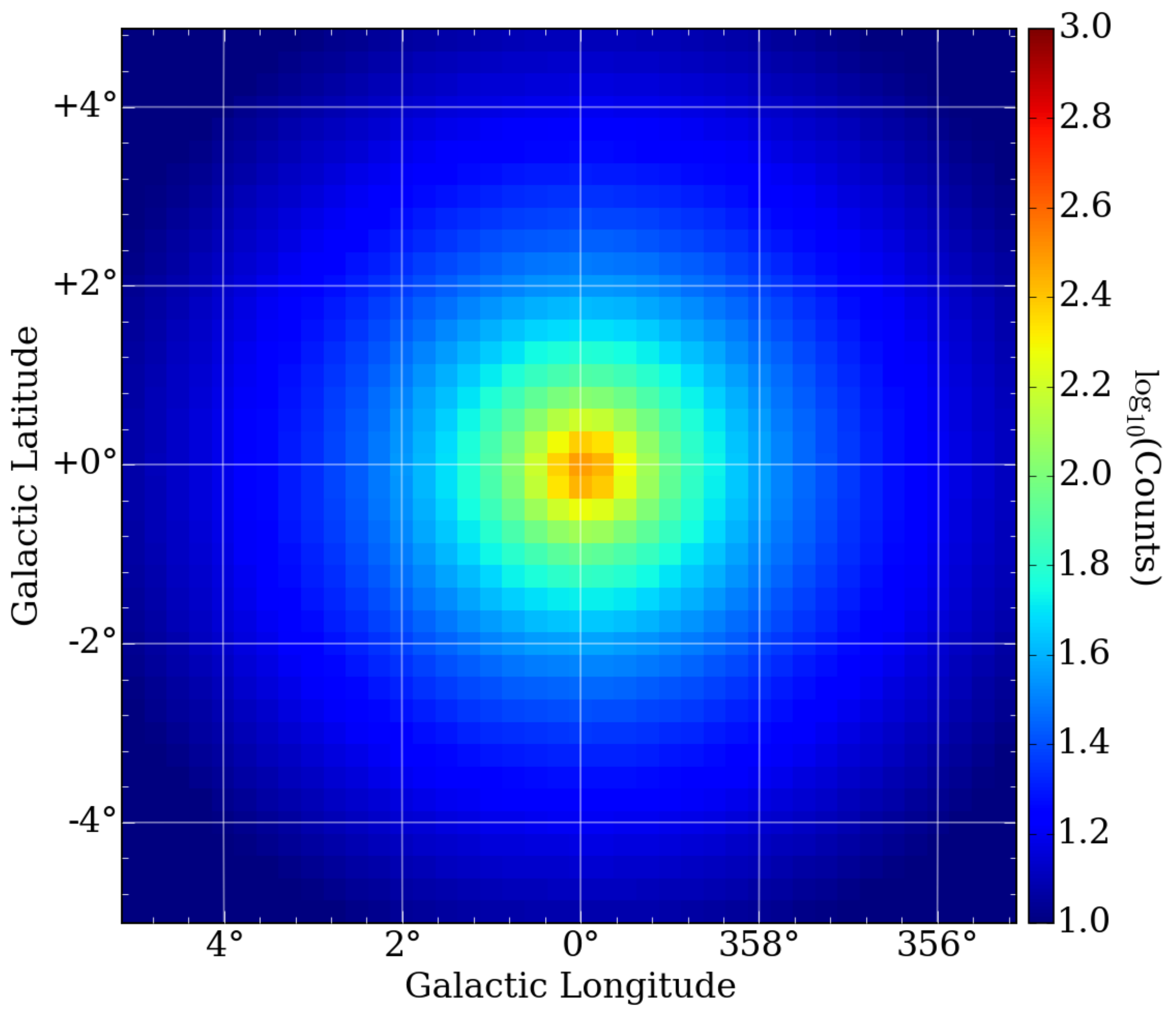}
\caption{{\it Left:} counts map of approximately 6 years of Fermi LAT gamma-ray data with energies from 1 to 35 GeV in the region +/- 5$^\circ$ from the Galactic Center, smoothed to 0.2$^\circ$ for visualization.  Much of the emission in this region is associated with the Galactic plane and is due to cosmic-ray interactions. Individual sources are also visible.  {\it Right:} model counts map of the same region for a dark matter annihilation signal which could explain the observed excess at GeV energies.
\label{fig:gevexcess}}  
\end{figure}

Several years ago an excess of gamma rays at GeV energies ($\sim 1$--10~GeV) from the Inner Galaxy was identified in the Fermi LAT data and has been confirmed by numerous subsequent studies~\cite{Goodenough:2009gk, Vitale:2009hr, Hooper:2010mq, Hooper:2011ti,Abazajian:2012pn,Gordon:2013vta, Hooper:2013rwa, Abazajian:2014fta, Daylan:2014rsa, TheFermi-LAT:2015kwa}.  This excess has been shown to be consistent with expectations for dark matter annihilation in both its angular distribution and its energy spectrum for $m_{\chi} \sim 10$--50~GeV, depending on the dominant final state, with an annihilation cross section within a factor of a few of thermal.  These detections were all obtained by modeling the Galactic diffuse emission and detected point sources, and then searching for an excess above the model (see Fig.~\ref{fig:gevexcess}).  Given the large uncertainties in the expected astrophysical diffuse emission, there has been debate over whether the excess represents a genuinely new component of Galactic gamma-ray emission or is an artifact of deficiencies in the diffuse model (see, e.g., \cite{Zhou:2014lva, Calore:2014nla, Gaggero:2015nsa}).  It has been suggested that the excess could potentially be explained by inverse Compton scattering by high-energy electrons injected in the past during a burst of activity at the Galactic Center such as accretion and star formation~\cite{Petrovic:2014uda}.  A similar argument has been made that an injection of cosmic-ray protons in the past could possibly lead to the observed excess~\cite{Carlson:2014cwa}.

The background models adopted in these analyses do not explicitly consider the contribution of unresolved members of known gamma-ray source populations, such as pulsars.  These sources are guaranteed to contribute to the total observed emission, however estimating their contribution requires extrapolation based on relatively few detected sources~\cite{Strong:2006hf}.  In particular, gamma-ray millisecond pulsars (MSPs) have been proposed as a possible explanation for the GeV excess~\cite{Abazajian:2010zy}.  MSPs have gamma-ray spectra that are reasonably consistent with the spectrum of the excess, and due to the evolution of these objects it is expected that they might be concentrated near the Galactic Center but also extend to high Galactic latitudes, consistent with the angular distribution of the excess.  There is not yet consensus on the size of the contribution of MSPs or other sources to the diffuse emission in the Inner Galaxy~\cite{Mirabal:2013rba, Hooper:2013nhl, Yuan:2014rca, Cholis:2014lta, Petrovic:2014xra}.  While analyses sensitive to sub-threshold sources suggest that such sources could provide an important contribution to the emission~\cite{Lee:2015fea, Bartels:2015aea}, curiously it does not appear that the sub-threshold sources are associated with known radio pulsars~\cite{Linden:2015qha}.

Secondary emission plays an important role in the interpretation of excess emission in this energy range due to the many processes that can contribute.  Bounds on a dark matter origin have been derived using multi-wavelength observations~\cite{Bringmann:2014lpa, Egorov:2015eta} and cosmic-ray data~\cite{Bringmann:2014lpa, Cirelli:2014lwa}.

Any resolution to the GeV excess will require new models of gamma-ray emission in the Inner Galaxy, and may lead to insights about the Galactic environment and the sources of cosmic rays and gamma rays.  Conclusive identification of the GeV excess as originating from dark matter annihilation will likely require a consistent detection in another target, in addition to a consistent multi-wavelength and multi-messenger picture from the Inner Galaxy.  In the meantime, the GeV excess remains an intriguing possible signal of dark matter annihilation.

\subsubsection{The 130 GeV line}
\label{sec:130gev}

In 2012 indications of hard spectral features were found in an analysis of the Fermi LAT data from the Galactic Center~\cite{Bringmann:2012vr}.  The signal was subsequently claimed to be a tentative detection of a line at $\sim 130$~GeV~\cite{Weniger:2012tx} -- a smoking gun dark matter signal.  This cautious discovery sparked a flurry of data analyses focused on further characterizing the signal~(e.g., \cite{Su:2012ft, Whiteson:2012hr}) and searching for it in other targets~(e.g,~\cite{GeringerSameth:2012sr,Hektor:2012kc}).  It was quickly recognized that the amplitude of the line signal was anomalously large given the lack of an associated continuum, which was challenging to accommodate in plain vanilla dark matter models~\cite{Cohen:2012me,Buchmuller:2012rc,Bai:2012qy}.  While the initial signal was confirmed by multiple analyses, later improvements to the analysis such as using updated Fermi LAT instrument response functions and incorporating the quality of the energy reconstruction of each event into the likelihood analysis failed to confirm a significant detection~\cite{Ackermann:2013uma}, and additional data has not increased the significance of the feature.  If the line feature is real, future gamma-ray telescopes, including H.E.S.S.-II, CTA, and the proposed GAMMA-400 telescope, should have the capabilities to conclusively detect it~\cite{Bergstrom:2012vd}.

\subsection{The Milky Way halo}

Moving away from the Galactic Center and looking out into the Milky Way halo has the advantage of substantially reducing backgrounds that concentrate near the Galactic Center and Galactic Plane~(e.g.,~\cite{Stoehr:2003hf, Serpico:2008ga}).  Selecting regions far from the Galactic Center also minimizes uncertainties in the expected flux due to lack of knowledge of the inner dark matter density profile, leading to more robust results.  Constraints on dark matter annihilation and decay have been placed based on Fermi LAT observations of diffuse emission at Galactic latitudes several degrees away from the Galactic Center~\cite{Cirelli:2009dv, Ackermann:2012rg}.

\subsubsection{The WMAP/Planck Haze}

Dark matter annihilation or decay in the Inner Galaxy is expected to generate substantial lower energy emission.  Due to propagation of charged cosmic rays, the secondary emission can extend far from the injection site of the cosmic rays, with synchrotron emission from dark matter at microwave frequencies expected to reach tens of degrees from the Galactic Center.

In 2004, Ref.~\cite{Finkbeiner:2003im} identified an excess of diffuse microwave emission from the Inner Galaxy in the WMAP data (the WMAP Haze).  Synchrotron emission from electrons injected by dark matter annihilation in the Milky Way was suggested by Ref.~\cite{Hooper:2007kb} as a possible origin of this signal, and those authors showed that the signal could be fit with dark matter particle parameters (mass, annihilation cross section, and density profile) consistent with expectations for WIMP cold dark matter.

The dark matter interpretation of the WMAP Haze implied that a corresponding gamma-ray signal should be observed due to the same population of electrons that produce the WMAP Haze inverse Compton scattering starlight up to gamma-ray energies.  Interestingly, a gamma-ray haze was soon identified in the Fermi LAT data~\cite{Dobler:2009xz}.  Further work, however, resolved the Fermi gamma-ray haze into distinct structures, which came to be called the Fermi Bubbles~\cite{Su:2010qj}.  The morphology of the Bubbles, in particular the observation of a hard edge, disfavors a dark matter interpretation; several astrophysical scenarios have been proposed.  

Recent analysis by the Planck collaboration has identified a microwave haze analogous to that found in the WMAP data~\cite{:2012rta}.  With the improved data set, morphological features in the Planck Haze  have been detected which strongly point to a common origin with the Fermi Bubbles, in particular a sharp edge in the microwave Haze aligned with the edge observed in the Fermi Bubbles.  However, it is still possible that some fraction of the Haze does have a dark matter origin, which could be the microwave counterpart of the GeV excess in the Inner Galaxy~\cite{Egorov:2015eta}.

\subsection{Subhalos}

Substructure in the Milky Way halo plays an important role in indirect searches.  Subhalos may be detected as individual sources in gamma-ray searches, while the collective emission from undetected subhalos contributes to diffuse emission at all frequencies.

The possibility of detecting individual subhalos in gamma rays was recognized very early~\cite{Lake:1990du}.  As numerical simulations achieved higher resolution and better characterized the subhalo population of a Milky-Way--like galaxy, more refined calculations of the gamma-ray signatures of substructure became possible~\cite{Diemand:2006ik, Pieri:2009je}.  Several studies have evaluated the implications of the abundance and properties of unidentified gamma-ray sources for dark matter models~(e.g., \cite{FlixMolina:2004ks, SiegalGaskins:2007dx, Brun:2010ci, Anderson:2010df, Belikov:2011pu, Zechlin:2012by, Mirabal:2012em, Ackermann:2012nb, Bertoni:2015mla, Schoonenberg:2016aml}), and constraints on dark matter properties have been placed based on the lack of suitable subhalo candidates among the unidentified sources.  The secondary emission at microwave frequencies from annihilation in subhalos may also be detectable over the CMB~\cite{Blasi:2002ct}.  Important uncertainties in these searches are in the properties of the subhalo population, including the subhalo mass function, density profiles, and radial distribution in the Galaxy.

The collective emission from unresolved subhalos contributes to the observed diffuse emission in all directions, and can strongly enhance the dark matter annihilation signal.  The emission from substructure is a small fraction of the total dark matter emission in the direction of the Galactic Center, but significantly brightens the dark matter signal in directions far away from the Inner Galaxy.  This enhancement to the dark matter signal, both in the Milky Way and in other dark matter search targets, is often referred to as a ``boost factor'', although the term is ambiguous, since it is also used to refer to other ways in which the signal is increased, such as through Sommerfeld enhancement.

The calculation of the emission from unresolved subhalos depends on the assumed characteristics of substructure.  The calculation is made more challenging by the need to extrapolate subhalo properties to mass scales 10 or more orders of magnitude smaller than those resolved in even the highest resolution numerical simulations.  As a result, predictions for the emission from substructure can vary wildly.  In recent years new approaches have been implemented to calculate the enhancement from substructure, and have somewhat tempered the trend towards very large boosts~\cite{Kamionkowski:2010mi, Sanchez-Conde:2013yxa}.  A study of CDM microhalos in numerical simulations also found that it is unlikely that the luminosity enhancement due to substructure in indirect search targets is larger than a factor of $\sim 10$~\cite{Anderhalden:2013wd}.  These studies have helped to reduce uncertainties in the  signals from dark matter substructure.
 
\subsection{Dark matter spikes around black holes}
 
The accumulation of dark matter around intermediate mass black holes (IMBHs) or supermassive black holes (SMBHs) may lead to detectable annihilation signals (e.g., \cite{Bertone:2009kj, Lacroix:2013qka, Belikov:2013nca}).  The dark matter distribution is expected to be extremely cuspy around these objects (forming a ``spike''), making them bright sources of gamma rays and other annihilation products.  Spikes around black holes may be detectable in individual objects or through their contribution to the intensity of the gamma-ray background. Moreover, spikes around IMBHs may be detectable through their contribution to the anisotropy of the gamma-ray background~\cite{Taoso:2008qz}.

\subsection{Milky Way satellite galaxies}

Dwarf galaxies and other known Milky Way satellites are some of the most dark-matter--dominated objects in the Universe.  In addition, due to these objects hosting few stars and little gas, astrophysical gamma-ray emission is expected to be negligible, making them very clean targets for indirect searches in gamma rays.

Many Milky Way satellites (dwarf galaxies and low-surface-brightness galaxies) have been evaluated as targets for gamma-ray searches~(e.g., \cite{Baltz:1999ra, Tyler:2002ux, Martinez:2009jh, Essig:2010em}), and several have been observed by  MAGIC~\cite{Aleksic:2013xea} and VERITAS~\cite{Acciari:2010ab, Aliu:2012ga}.  No excesses were observed by these IACTs, and limits on dark matter fluxes were placed.  The Fermi LAT sky survey observing mode allows many satellite galaxies to be used as targets, enabling joint likelihood analyses of multiple sources to improve sensitivity.  Null searches with Fermi LAT data considering multiple targets have yielded some of the strongest constraints on the dark matter annihilation cross section to date~\cite{GeringerSameth:2011iw, Ackermann:2011wa, Ackermann:2013yva, Ackermann:2015zua}.

At a distance of only 50~kpc with a halo mass of~$\sim 10^{10}$~M$_{\odot}$ and as the largest known satellite of the Milky Way, the Large Magellanic Cloud (LMC) is an obvious candidate for indirect searches.  Emission associated with dark matter annihilation in the LMC has been considered at both gamma-ray~\cite{Gondolo:1993yj, Tasitsiomi:2003vw, Fornengo:2004kj} and radio frequencies~\cite{Tasitsiomi:2003vw, Siffert:2010cc}.  Recently a search for dark matter annihilation in the LMC was performed with the Fermi LAT data, and no significant signal was found~\cite{Buckley:2015doa}.  A study following a similar approach but focused on the Small Magellanic Cloud likewise yielded a null result and set bounds on the dark matter emission~\cite{Caputo:2016ryl}.

\subsection{External galaxies}

Nearby external galaxies can be excellent targets for indirect dark matter searches.  Compared to satellite galaxies, these are typically more massive, which can compensate in the J-factor for their larger distance.  External galaxies are often well-characterized using multi-wavelength data, and, for some instruments such as IACTs, being able to contain the object within the field of view can be an advantage over signals from the Milky Way that are very extended.

The Andromeda Galaxy (M31) is the closest Milky-Way--like galaxy, at $\sim 800$~Mpc with a halo mass of~$\sim 10^{12}$~M$_{\odot}$, comparable to the mass of the Milky Way.  It has been considered in both gamma rays~\cite{Fornengo:2004kj, Mack:2008wu, Dugger:2010ys} and radio~\cite{Egorov:2013exa}.  M31 has a sizable astrophysical background due to cosmic-ray interactions and unresolved source populations, but its location far from the Milky Way's Galactic Center and associated backgrounds is advantageous~\cite{Fornengo:2004kj}.  Radio observations are found to be the most constraining under the assumption of a relatively strong magnetic field~\cite{Egorov:2013exa}.  A recent analysis of Fermi LAT observations of M31~\cite{Li:2013qya} placed constraints on dark matter annihilation comparable to those from Fermi LAT observations of dwarf galaxies.

Other nearby galaxies have also been considered for indirect detection, including M87~\cite{Baltz:1999ra, Fornengo:2004kj, Saxena:2011tk} and M33~\cite{Borriello:2009tt}.

\subsection{Galaxy clusters}

Nearby galaxy clusters such as Virgo, Fornax, and Coma are interesting targets for dark matter searches.  In optimistic scenarios for annihilation, clusters can be competitive, however recent estimates of the dark matter flux based on new approaches to substructure modeling make a less compelling case for these targets.  Constraints on annihilation and decay have been derived from null searches in Fermi LAT data~\cite{Ackermann:2010rg, Dugger:2010ys, Huang:2011xr, Cirelli:2012ut, Ackermann:2015fdi}.

\subsection{Cosmological fluxes}
\label{sec:cosmophotons}

The annihilation or decay of dark matter particles throughout the universe produces large fluxes of diffuse emission spanning a range of wavelengths with spectral and spatial signatures that can help to distinguish this signal from other non-exotic astrophysical sources.  Spectral signatures arise both from secondary interactions and from redshifting~\cite{Bergstrom:2001jj}, while the clumpy distribution of dark matter in halos gives rise to anisotropies in diffuse emission.  

\subsubsection{Prompt radiation and the gamma-ray background}

The prompt emission from cosmological WIMP annihilation and decay is observed as a contribution to the isotropic gamma-ray background (IGRB), the statistically-isotropic observed all-sky diffuse emission at gamma-ray energies.  I refer the reader to~\cite{Fornasa:2015qua} for a comprehensive review of the current status of the measurement and interpretation of the IGRB\@.  A presentation of the calculation of observed gamma-ray emission from cosmological dark matter annihilation is given in~\cite{Ullio:2002pj} and can be adapted for decay scenarios.   A power-spectrum based approach to the calculation is presented in~\cite{Serpico:2011in,Sefusatti:2014vha}.  The observed gamma-ray spectrum is distorted due to EBL absorption and redshifting, resulting in line emission being observed as asymmetrically-broadened peaks, and a shift of the peak of the continuum spectrum to lower energies.  

Galactic dark matter, both in the smooth halo and in substructure, contributes to the observed IGRB since its angular distribution on the sky is approximately isotropic on large angular scales, and in practice the gradient in the emission towards the Galactic Center is difficult to disentangle from the large astrophysical backgrounds in that direction.  Any detections or constraints based on measurements of the properties of the IGRB must therefore consider contributions from both Galactic and extragalactic dark matter.

The Fermi LAT measurement of the IGRB spectrum allows strong constraints to be placed on dark matter fluxes from annihilation and decay in certain scenarios~\cite{Abdo:2010dk, Abazajian:2010zb, Cirelli:2012ut}.  While conservative constraints can be obtained by allowing dark matter to account for the entirety of the observed IGRB, more realistic scenarios can be considered by modeling the astrophysical populations which are expected to contribute the bulk of the IGRB intensity~\cite{Calore:2013yia, Ackermann:2015tah}.  These populations include starforming galaxies, blazars, and misaligned active galactic nuclei.

The anisotropy of the IGRB is a complementary observable that can be used to constrain the properties of contributing source populations, including dark matter.  The IGRB is expected to exhibit anisotropies since it originates from unresolved sources.  The dark matter contribution to the anisotropy has been considered in several studies.  In most scenarios the Galactic dark matter emission from subhalos is characterized by a large fractional anisotropy which may allow it to be detectable even over a large background intensity~\cite{SiegalGaskins:2008ge, Fornasa:2009qh, SiegalGaskins:2009ux, Hensley:2009gh, Ando:2009fp, Fornasa:2012gu, Calore:2014hna}.  The anisotropy from extragalactic dark matter is generally smaller than from Galactic dark matter~\cite{Ando:2005xg, Ando:2006cr, Fornasa:2009qh,  Fornasa:2012gu}, however variations in substructure models can affect this comparison.

In 2012 the Fermi LAT collaboration measured the anisotropy of the IGRB for the first time~\cite{Ackermann:2012uf}, enabling new constraints to be placed on the contribution of dark matter~\cite{Ackermann:2012uf, Ando:2013ff, Gomez-Vargas:2014yla} and other source populations~\cite{Ackermann:2012uf, Cuoco:2012yf, Harding:2012gk, DiMauro:2014wha} to the IGRB based on requiring that the predicted anisotropy not exceed that measured in the data.  As blazars are expected to contribute the majority of the IGRB anisotropy, robust predictions for the amplitude of their anisotropy can  enhance sensitivity to dark matter signals by limiting the anisotropy available to be attributed to dark matter.    Another approach taking advantage of angular information in diffuse backgrounds is to cross-correlate the emission with catalogs or other tracers of the dark matter distribution~\cite{Xia:2011ax, Camera:2012cj, Fornengo:2013rga, Ando:2013xwa, Shirasaki:2014noa}.  The use of the 1-pt PDF to detect the contribution of dark matter to the gamma-ray background via clustering has also been proposed~\cite{Lee:2008fm, Dodelson:2009ih, Baxter:2010fr}.

\subsubsection{Multi-wavelength radiation}
\label{sec:cosmomulti}

Multi-wavelength observations can play an important role in indirect dark matter searches in all-sky diffuse emission since secondary emission is relevant for both cosmological and Galactic dark matter.

Inverse Compton scattering of CMB photons by electrons produced by dark matter annihilation at all redshifts can lead to large contributions to the gamma-ray and X-ray diffuse backgrounds.  Leptophilic dark matter, which annihilates or decays primarily to charged leptons, has garnered interest because it could explain the PAMELA positron fraction measurement as well as other cosmic-ray data (see~\S\ref{sec:positron}), however in many cases the models that can explain the PAMELA data are incompatible with the measured gamma-ray and/or X-ray backgrounds~\cite{Profumo:2009uf,Belikov:2009cx,Huetsi:2009ex}.  Large rates of annihilation to leptons in Galactic dark matter substructure can also lead to tension with measured gamma-ray backgrounds~\cite{Kistler:2009xf}.

The cross-correlation approach mentioned in the previous section has not only been used with the IGRB, but has also been considered for secondary emission from dark matter annihilation and decay at lower frequencies including X-ray and radio~\cite{Fornengo:2013rga}, and found to be a promising tool for dark matter detection.

Interesting for multi-wavelength dark matter searches, an excess in the radio background has been measured by the ARCADE~2 experiment~\cite{Fixsen:2009xn,Seiffert:2009xs}.  The measured flux is roughly a factor of 5 larger than expected from extrapolations of known source populations and currently there are only upper limits on its anisotropy; a consistent model for this emission requires that the sources are extremely numerous and faint~\cite{Holder:2012nm, Singal:2009dv}.  One possible explanation is synchrotron emission from cosmological dark matter annihilation~\cite{Fornengo:2011cn, Hooper:2012jc}.

\section{Local cosmic-ray fluxes from WIMP annihilation and decay}
\label{sec:wimpcr}

Measurements of cosmic-ray fluxes at Earth are a unique probe of local dark matter annihilation and decay.  Cosmic-ray observations provide complementary information to that from photon and neutrino searches, and can place strong constraints on branching ratios to specific channels.  Observations of apparent anomalies in local cosmic-ray fluxes that could be explained by dark matter have led to much interest.  Two such anomalies are discussed below -- the rising positron fraction and the cosmic-ray electron+positron excess -- along with the sensitivity to dark matter from anti-protons and heavier nuclei.

\subsection{The rising positron fraction and the electron+positron excess}
\label{sec:positron}

Several experiments have observed that the positron fraction (number of positrons / number of electrons + positrons) rises from $\sim 10$~GeV to at least $\sim 100$~GeV.  While earlier experiments showed indications of this feature, it was seen clearly in the PAMELA data~\cite{Adriani:2008zr}, and subsequently confirmed by the Fermi LAT~\cite{FermiLAT:2011ab} and most recently by AMS-02~\cite{Aguilar:2013qda}.

The rise in the positron fraction contradicts the conventional expectation for Galactic cosmic rays from secondary production, which instead predicts a fall in the positron fraction above $\sim 10$~GeV.  
The observed rise suggests that there must be a nearby source injecting positrons at high energies. Dark matter annihilation or decay is one possibility, although pulsars have also been proposed as the origin (e.g.,~\cite{Profumo:2008ms}).  Both dark matter and pulsars are expected to inject electron-positron pairs, and are sufficiently close to the Earth to explain the observed high-energy positrons.   Analysis of the AMS-02 results suggests that in principle both explanations are viable from the perspective of generating the observed fluxes~\cite{Yuan:2013eja}.

After the PAMELA measurement was reported, another apparent cosmic-ray anomaly was found by the ATIC collaboration.  ATIC measured the total electron+positron spectrum, and discovered an excess at energies of several hundred GeV over the conventional background model~\cite{Chang:2008aa}.  Simultaneous explanations for this and the positron fraction were proposed, with dark matter as one compelling possibility (e.g.,~\cite{Cholis:2008wq}).  A later measurement of the electron+positron spectrum by the Fermi LAT did not confirm the large feature observed by ATIC, however it found a small but noticeable excess at a few hundred GeV~\cite{Ackermann:2010ij,Abdo:2009zk}.  H.E.S.S. also did not confirm the ATIC peak in the measured electron+positron spectrum, but reported a steepening in the spectrum above $\sim 1$~TeV~\cite{Aharonian:2009ah}.

A dark matter interpretation of the positron fraction and ATIC/Fermi excess generically requires large annihilation cross sections compared to the thermal relic value, and leptophilic dark matter.  This is necessary both to generate the required flux of electrons/positrons, and to avoid constraints from overproducing anti-protons.  Numerous multi-messenger constraints on this scenario have been identified, including associated gamma-ray fluxes from inverse Compton scattering that are in tension with observational bounds~\cite{Zhang:2008tb, Kistler:2009xf, Cirelli:2009vg, Pato:2009fn, Borriello:2009fa, Abdo:2010dk, Slatyer:2011kg, Ackermann:2012rg} and distortions of the CMB~\cite{Galli:2009zc,Slatyer:2009yq}, making it challenging to construct a dark matter model that explains all of the cosmic-ray data while being compatible with other data sets. 

Both the amplitude of the positron fraction and also its spectrum may provide a means of distinguishing the origin of the emission and constraining dark matter interpretations~\cite{Bergstrom:2008gr, Cirelli:2008pk, Malyshev:2009tw, Bergstrom:2013jra, Cholis:2013psa}.  A dark matter origin implies a cut-off at the dark matter particle mass (or half the mass in the case of decay), while if multiple pulsars are the source of the additional positrons, several spectral features would be expected due to the varying cut-offs in the spectra of different pulsars.  

Another handle for identifying the origin of the positron excess is an anisotropy in the arrival directions of the positrons or electrons+positrons~\cite{Hooper:2008kg, Cernuda:2009kk, DiBernardo:2010is, Linden:2013mqa}.  If a single nearby source generated the rise in the positron fraction, despite significant loss of directional information due to diffusion, a small anisotropy should remain in the angular distribution of the positrons and electrons+positrons.  Even in the case of multiple sources such an anisotropy might appear, dominated by the nearest, strongest source.  In general, the smooth dark matter distribution should generate only a very small anisotropy in the direction of the Galactic Center, while a pulsar-induced anisotropy could be much stronger and in any direction; a nearby dark matter subhalo large and close enough to produce the positron excess is excluded by the non-observation of gamma rays from it~\cite{Profumo:2014yxa}.  Currently there are only upper limits on the anisotropy of the cosmic-ray electron+positron flux from the Fermi LAT~\cite{Ackermann:2010ip} and on the anisotropy of the positron fraction from AMS-02~\cite{Aguilar:2013qda}, and to date they are not sufficiently strong to exclude proposed dark matter or pulsar scenarios.

\subsection{Constraints from anti-protons and other nuclei}

Final states that yield hadronic products contribute to the production of anti-proton cosmic rays, thus the measurement of an excess of anti-protons could provide a unique and sensitive signature of dark matter annihilation or decay.  Current measurements of the anti-proton flux yield strong bounds on dark matter models~\cite{Pato:2009fn, Evoli:2011id, Cirelli:2013hv, Fornengo:2013xda}, and have been used to test dark matter interpretations of the Galactic Center gamma-ray excess~\cite{Cirelli:2014lwa, Bringmann:2014lpa}.  Other cosmic-ray abundances also have implications for interpretations of cosmic-ray anomalies, e.g., the measured boron-to-carbon ratio constrains scenarios in which the observed rise in the positron fraction is due to acceleration in nearby supernova remnants~\cite{Cholis:2013lwa}.

Heavier anti-nuclei provide important sensitivity as well.  Anti-deuterons have long been recognized as a promising cosmic-ray channel due to the lack of astrophysical backgrounds~\cite{Donato:1999gy, Kadastik:2009ts, Dal:2012my, Hailey:2013gwa, Fornengo:2013osa}; AMS-02 and GAPS are expected to have interesting sensitivity to anti-deuterons from dark matter annihilation and decay.  Anti-helium has also been identified as a complementary channel for detection~\cite{Carlson:2014ssa,Cirelli:2014qia}.  The backgrounds are extremely small, however the flux of anti-helium from dark matter is predicted unfortunately to be too low to be detectable with AMS-02 or GAPS.

\section{Neutrinos from WIMP annihilation and decay}
\label{sec:wimpnu}

Neutrinos provide a complementary channel for detecting dark matter annihilation and decay.  Indirect searches with neutrinos can access many of the same targets as photon-based searches, with the Galactic Center and the Milky Way halo two of the most favorable.  Neutrino telescopes can also search for neutrinos from annihilation of WIMP dark matter particles in the Sun and Earth, as described below.

\subsection{The Galactic Center and Milky Way Halo}

IceCube~\cite{Abbasi:2012ws} and Super-K~\cite{Desai:2004pq} have published constraints on WIMP fluxes from null searches for prompt neutrinos from dark matter annihilation in the Galactic Center, and IceCube has also used two different approaches to search for prompt neutrinos from WIMPs in the Milky Way halo~\cite{Abbasi:2011eq,Aartsen:2014hva}.  The IceCube analyses yielded limits on dark matter annihilation and decay for several channels for $m_{\chi} \sim 100$~GeV -- 10~TeV.  Constraints on hard channels are stronger than those on soft channels by several orders of magnitude.  The Super-K Galactic Center analysis placed upper limits on the WIMP-induced neutrino flux for $m_{\chi} = 20$~GeV -- 10~TeV.

\subsection{Neutrinos from the Sun and Earth}

WIMPs gravitationally captured by scattering interactions with nucleons in the Sun and Earth could then annihilate and produce a detectable neutrino flux~\cite{Gould:1987ir, 1992ApJ...388..338G, Wikstrom:2009kw}.  Once captured, a WIMP continues to scatter, losing energy and sinking to the core of the Sun or Earth; a large abundance of WIMPs can accumulate through this mechanism.  At the core the WIMPs annihilate, but only neutrinos can escape.  For standard parameters, the abundance of WIMPs captured via scattering is large enough that capture and annihilation have reached equilibrium in the Sun (i.e., the rate of capture is twice the rate of annihilation), but not in the Earth.  In the equilibrium case, the annihilation flux depends on the WIMP-nucleon scattering cross section, which determines the capture rate, and is independent of the annihilation cross section, allowing this indirect search technique to probe the same parameter space as laboratory direct-detection experiments.   Bounds on the scattering cross section as a function of WIMP mass have been placed via null searches for an excess from the direction of the Sun by IceCube~\cite{Aartsen:2012kia}, ANTARES~\cite{Adrian-Martinez:2013ayv}, and Super-K~\cite{Tanaka:2011uf}.  The Sun offers particularly strong sensitivity to spin-dependent scattering, as the Sun is primarily a hydrogen target.  In addition, astrophysical uncertainties in the capture rate of WIMPs by the Sun are small, making limits from neutrino searches robust.  Super-K also performed a search for an excess of neutrinos from the Earth~\cite{Desai:2004pq}.  These searches all focused on prompt neutrinos produced in annihilation with $\sim$~GeV and greater energies.

Low-energy ($\sim$~tens of MeV) neutrinos have been proposed as another channel for detecting WIMP annihilation in the Sun~\cite{Bernal:2012qh,Rott:2012qb}.  This approach notes that all SM final states with hadronic decay modes produce pions, but, unlike in vacuum where the pions quickly decay, in the dense Solar medium the pions lose energy by scattering off nucleons.  Each scattering interaction produces more pions, with a large fraction of the energy in hadronic products ultimately ending up as pions.  The positive pions decay at rest, producing 3 neutrinos with energies of $\sim 20$--55~MeV with known spectra.  Refs.~\cite{Bernal:2012qh,Rott:2012qb} proposed to search for this signature in Super-K data, and demonstrated that this channel is complementary to and competitive with high-energy neutrino searches and direct detection experiments.

Dark matter capture rates in the Sun and Earth and associated neutrino spectra, including low-energy neutrinos produced through subsequent interactions, are calculated in~\cite{Baratella:2013fya} and available in numerical form\footnote{\tt http://www.marcocirelli.net/PPPC4DMID.html}.

\section{Indirect signals from superheavy dark matter decay}
\label{sec:superheavy}

If dark matter is a superheavy particle, its decay or annihilation to SM particles could be detected using UHECR detectors such as the Pierre Auger Observatory.  The Greisen-Zatsepin-Kuzmin (GZK) limit restricts the energy of cosmic rays propagating over cosmological distances to be below~$\sim 5 \times 10^{19}$~eV due to interactions of ultra-high-energy (UHE) protons with the CMB that cause energy loss via pion production through the delta resonance~\cite{Greisen:1966jv,Zatsepin:1966jv}.  It has been proposed that decay of nearby superheavy dark matter particles could account for some or all of the measured UHECR flux above the GZK bound~\cite{Berezinsky:1997hy, Blasi:2001hr}.  A distinguishing signature of this scenario would be an anisotropy in the arrival directions of UHECRs, due to the anisotropic distribution of dark matter in the Milky Way halo with respect to Earth.  Other nearby targets, such as the Virgo cluster, may also lead to an overdensity of UHECRs from a particular sky direction~\cite{Berezinsky:1997hy}.

Gamma rays may also be detected from superheavy dark matter annihilation or decay.  The gamma-ray flux is the sum of the prompt gamma-ray emission and secondary emission depending on the ambient environment and the final state.  For Galactic dark matter, secondary emission can result from electrons and positrons emitting synchrotron at gamma-ray energies due to propagation in the Galactic magnetic field~\cite{Blasi:1999nr}.  For cosmological dark matter, prompt high-energy gamma rays pair produce with the EBL, leading to high-energy electron-positron pairs.  The subsequent interactions of the electrons and positrons (inverse Compton scattering of the EBL and synchrotron emission in the intergalactic magnetic field) result in a cascade of lower-energy gamma rays.  For superheavy dark matter models, these gamma rays may be detectable as a component of the IGRB~\cite{Murase:2012xs}.

UHE neutrinos would also be produced as prompt emission products for many final states, and do not suffer from GZK attenuation.  Limits on VHE neutrino fluxes from the Auger Observatory and from neutrino telescopes including AMANDA, ANITA, and IceCube place bounds on superheavy dark matter~\cite{Albuquerque:2000rk, Halzen:2001ec, Esmaili:2012us, Murase:2012xs}.  The recent observation of two PeV neutrinos by IceCube has been considered as a possible line emission signature of superheavy dark matter decay~\cite{Feldstein:2013kka}.

\section{Sterile neutrino decay signals}
\label{sec:sterilenu}

Indirect searches for line emission from sterile neutrino decay probe the parameter space of the sterile neutrino mass and its mixing angle with active neutrinos, which sets the rate of decay.  Null searches in X-rays and gamma rays have excluded regions at large mixing angles and large masses~\cite{Boyarsky:2005us, Boyarsky:2006ag, Abazajian:2006jc, Loewenstein:2008yi, Ng:2015gfa}, while complementary but model-dependent constraints have been derived from measurements of small-scale structure~\cite{Viel:2005qj, Boyarsky:2008ju} and cosmological lepton number, bounding the parameter space at small masses and small mixing angles, respectively.  Despite constraints from all sides, a window of parameter space remains for sterile neutrino masses of $\sim 10-50$~keV~\cite{Abazajian:2006yn}.

\subsection{The 3.5 keV line}

In recent years there have been claims of a detection of line emission at 3.5 keV in galaxy clusters~\cite{Bulbul:2014sua, Boyarsky:2014jta} and in M31~\cite{Boyarsky:2014jta}. There is no consensus on an astrophysical or instrumental explanation for the claimed line, which has been interpreted as a possible signature of the decay of 7~keV sterile neutrino dark matter.  No corresponding signal has been detected from the Milky Way~\cite{Riemer-Sorensen:2014yda} or from Draco~\cite{Jeltema:2015mee}, leading to some tension for a dark matter interpretation.  Future indirect searches using alternative targets may help to clarify the origin of the line feature and robustly test a dark matter interpretation.

\section{Summary}
\label{sec:disc}

\begin{figure}
\centering
\includegraphics[width=0.9\textwidth]{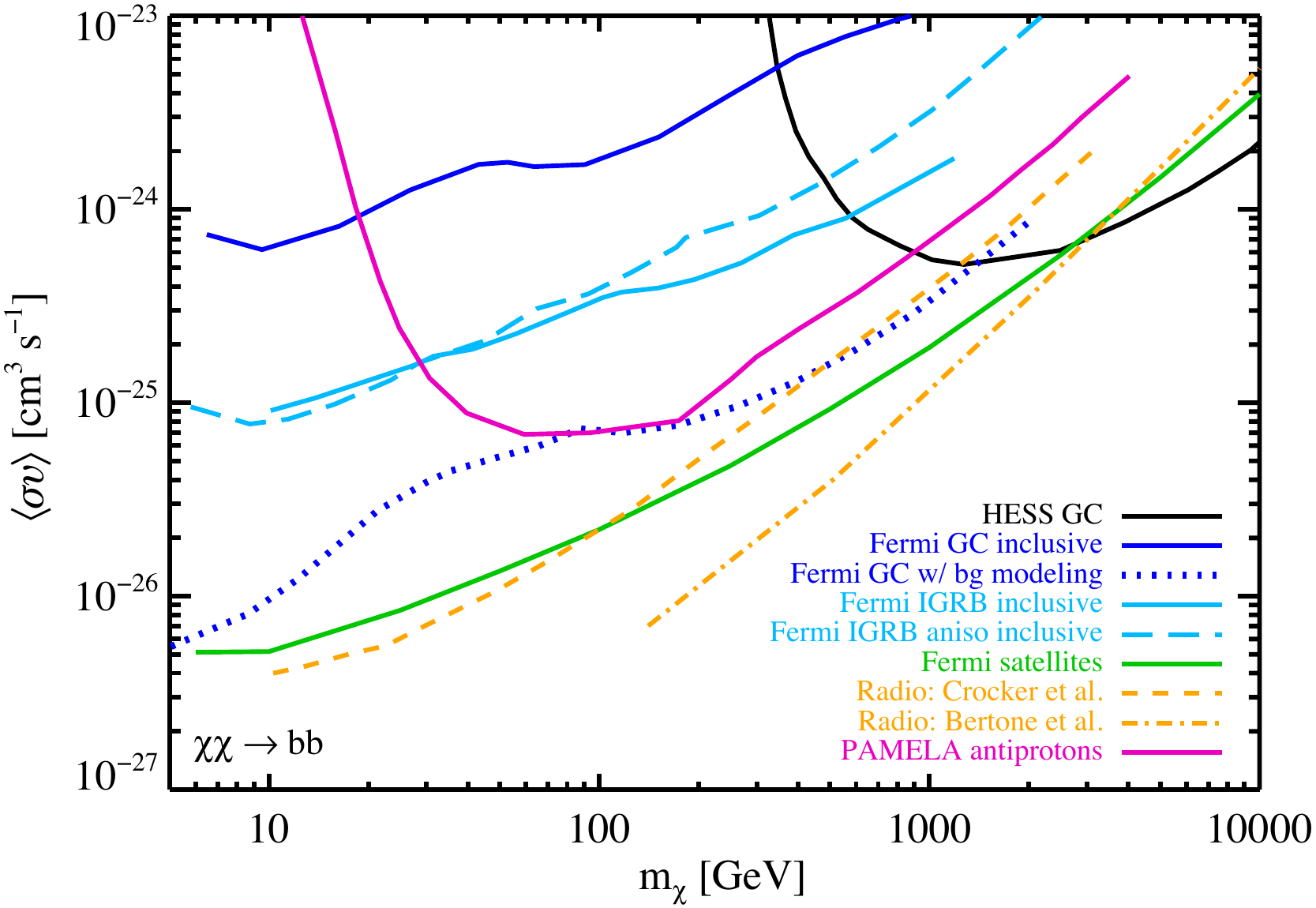}
\caption{Constraints on WIMP dark matter models from selected indirect search analyses for annihilation to $b\bar{b}$.  The parameter space above each curve is excluded at 95\% confidence level.  Gamma-ray constraints are shown from a H.E.S.S. analysis of the Galactic Center halo~\cite{Abramowski:2011hc}, and several analyses of Fermi LAT data: the Galactic Center without background modeling (``inclusive'')~\cite{Gomez-Vargas:2013bea} and with background modeling~\cite{Hooper:2012sr}, the  IGRB spectrum without background modeling~\cite{Abdo:2010dk}, the IGRB anisotropy without background modeling~\cite{Gomez-Vargas:2014yla}, and a combined analysis of Milky Way satellite galaxies~\cite{Ackermann:2015zua}.  Also shown are constraints obtained in Bertone et al.~2009~\cite{Bertone:2008xr} and Crocker et al.~2010~\cite{Crocker:2010gy} using radio observations of the Galactic Center, and constraints from the PAMELA anti-proton data from~\cite{Cirelli:2013hv}.  See text for details.\label{fig:constraints}}  
\end{figure}

As discussed in the previous sections, indirect detection of dark matter has been tackled with a variety of instruments and data sets.  To give a sense of the current status of indirect searches, a compilation of  selected constraints in the parameter space of the WIMP mass and annihilation cross section is shown in Fig.~\ref{fig:constraints} for annihilation to $b\bar{b}$.  For this final state, searches using positrons and neutrinos are not competitive and are not shown in the figure.  All curves represent the 95\% confidence level upper limits on the annihilation cross section.

At the very highest masses, the strongest bounds are from a H.E.S.S. analysis of the Galactic Center halo~\cite{Abramowski:2011hc}.  Limits from analyses with Fermi LAT data span the entire WIMP mass range shown.  Limits obtained without modeling non-exotic contributions and instead assuming the entire measurement is available to attribute to a dark matter signal are referred to as ``inclusive''; inclusive limits are shown for the Galactic Center~\cite{Gomez-Vargas:2013bea}, the IGRB intensity spectrum~\cite{Abdo:2010dk}, and the IGRB anisotropy~\cite{Gomez-Vargas:2014yla}.  These should be taken as quite robust upper limits.  The limits in these studies strengthen significantly when background modeling is performed.  As an example, bounds obtained from an analysis of the Galactic Center in which modeling of the background was performed are also shown~\cite{Hooper:2012sr}.  This limit is model-dependent, but somewhat more ``realistic'' since a large contribution to the total emission from non-exotic processes in the Galactic Center is guaranteed.  The large improvement in sensitivity indicates that a better understanding of backgrounds can play an important role in enhancing the prospects for indirect searches.  For the IGRB intensity spectrum, the curves shown are the ``conservative'' limits assuming the ``BulSub'' substructure model as described in~\cite{Abdo:2010dk}.  The IGRB anisotropy bounds are preliminary, and derived from the published anisotropy measurement~\cite{Ackermann:2012uf}.
The limits from a recent analysis of Fermi LAT observations of satellite galaxies are shown as well~\cite{Ackermann:2015zua}.  

Constraints from Bertone et al.~2009~\cite{Bertone:2008xr} and Crocker et al.~2010~\cite{Crocker:2010gy} using radio observations provide some of the strongest bounds on the annihilation cross section for this channel.  However, the precise limits depend sensitively on assumptions about the Galactic magnetic field profile, especially at small radii, which is poorly constrained, and thus these limits cannot be considered robust.  The limits from PAMELA antiproton data from~\cite{Cirelli:2013hv} are also dependent on the assumed propagation model.  An improved understanding of the Galactic environment could help make observations of secondary emission and local cosmic-ray measurements more robust indirect search tools.

Note that the results shown in Fig.~\ref{fig:constraints} for gamma rays and radio emission from the Galactic Center were obtained for a NFW profile, and the local antiproton flux results were obtained for an Einasto profile.  The analyses adopted slightly different choices for the halo parameters, including different values for the local dark matter density.  The expected shift in the limits if all analyses considered the same density profile is no more than a factor of~$\sim 4$.

Current limits are already beginning to constrain the canonical thermal relic annihilation cross section for WIMP dark matter, and the reach of upcoming experiments is expected to dramatically increase; it is clear that the next several years will be a key time for indirect searches.  With possible detections of different candidates already claimed in gamma rays and X-rays, and suggestive anomalies reported in cosmic rays and the radio background, sensitive indirect searches in multiple targets as well as multi-wavelength and multi-messenger searches will be necessary to confirm or invalidate these claims.  In the coming years there is the exciting potential to not only robustly detect dark matter through its annihilation or decay signals, but also to characterize its distribution, and pin down the fundamental properties of this mysterious particle.

\section{Acknowledgements}

It is a pleasure to thank Kevork Abazajian, Lars Bergstr\"{o}m, Gianfranco Bertone, Ilias Cholis, Marco Cirelli, J\"{u}rg Diemand, Nicolao Fornengo, Dan Hooper, Shunsaku Horiuchi, Stefano Profumo, Carsten Rott, and Gabrijela Zaharijas for helpful comments on this manuscript.  I acknowledge support from NASA through Einstein Postdoctoral Fellowship grant PF1-120089 awarded by the Chandra X-ray Center, which is operated by the Smithsonian Astrophysical Observatory for NASA under contract NAS8-03060.  I also thank the GRAPPA Institute at the University of Amsterdam and the Institut d'Astrophysique de Paris for hospitality during the final stages of this work.

\bibliography{indirect}

\label{lastpage}

\end{document}